\documentclass[twocolumn]{aastex631}
\usepackage[utf8]{inputenc}
\usepackage{soul}
\usepackage{amsmath}

\usepackage{dsfont}

\begin{document}

\title{The CEERS Photometric and Physical Parameter Catalog}

\author[0000-0002-1803-794X]{Isa G. Cox}
\affiliation{Laboratory for Multiwavelength Astrophysics, School of Physics and Astronomy, Rochester Institute of Technology, 84 Lomb Memorial Drive, Rochester, NY
14623, USA}
\email{igc5972@rit.edu}

\author[0000-0001-9187-3605]{Jeyhan S. Kartaltepe}
\affiliation{Laboratory for Multiwavelength Astrophysics, School of Physics and Astronomy, Rochester Institute of Technology, 84 Lomb Memorial Drive, Rochester, NY 14623, USA}

\author[0000-0002-9921-9218]{Micaela B. Bagley}
\affiliation{NASA-Goddard Space Flight Center, Code 665, Greenbelt, MD, 20771, USA}
\affiliation{Department of Astronomy, The University of Texas at Austin, Austin, TX, USA}

\author[0000-0001-8519-1130]{Steven L. Finkelstein}
\affiliation{Department of Astronomy, The University of Texas at Austin, Austin, TX, USA}
\affiliation{Cosmic Frontier Center, The University of Texas at Austin, Austin, TX, USA}

\author[0000-0002-8018-3219]{Caitlin Rose}
\affil{Laboratory for Multiwavelength Astrophysics, School of Physics and Astronomy, Rochester Institute of Technology, 84 Lomb Memorial Drive, Rochester, NY 14623, USA}

\author[0000-0002-0101-336X]{Ali Ahmad Khostovan}
\affiliation{Department of Physics and Astronomy, University of Kentucky, 505 Rose Street, Lexington, KY 40506, USA}
\affiliation{Laboratory for Multiwavelength Astrophysics, School of Physics and Astronomy, Rochester Institute of Technology, 84 Lomb Memorial Drive, Rochester, NY 14623, USA}

\author[0000-0003-4922-0613]{Katherine Chworowsky}\altaffiliation{NSF Graduate Fellow}
\affiliation{Department of Astronomy, The University of Texas at Austin, Austin, TX, USA}
\affiliation{Cosmic Frontier Center, The University of Texas at Austin, Austin, TX, USA}

\author[0000-0002-7303-4397]{Olivier Ilbert}
\affiliation{Aix Marseille Univ, CNRS, CNES, LAM, Marseille, France  }

\author[0000-0002-6610-2048]{Anton M. Koekemoer}
\affiliation{Space Telescope Science Institute, 3700 San Martin Dr., Baltimore, MD 21218, USA}

\author[0000-0001-7113-2738]{Henry C. Ferguson}
\affil{Space Telescope Science Institute, 3700 San Martin Dr., Baltimore, MD 21218, USA}

\author[0000-0002-7959-8783]{Pablo Arrabal Haro}
\altaffiliation{NASA Postdoctoral Fellow}
\affiliation{Astrophysics Science Division, NASA Goddard Space Flight Center, 8800 Greenbelt Rd, Greenbelt, MD 20771, USA}

\author[0000-0001-8534-7502]{Bren E. Backhaus}
\affiliation{Department of Physics and Astronomy, University of Kansas, Lawrence, KS 66045, USA}

\author[0000-0001-5414-5131]{Mark Dickinson}
\affiliation{NSF's National Optical-Infrared Astronomy Research Laboratory, 950 N. Cherry Ave., Tucson, AZ 85719, USA}

\author[0000-0003-3820-2823]{Adriano Fontana}
\affiliation{INAF - Osservatorio Astronomico di Roma, via di Frascati 33, 00078 Monte Porzio Catone, Italy}

\author[0000-0002-4162-6523]{Yuchen Guo}
\affiliation{Department of Astronomy, The University of Texas at Austin, Austin, TX, USA}
\affiliation{Cosmic Frontier Center, The University of Texas at Austin, Austin, TX, USA}

\author[0000-0002-5688-0663]{Andrea Grazian}
\affiliation{INAF--Osservatorio Astronomico di Padova, Vicolo dell'Osservatorio 5, I-35122, Padova, Italy}

\author[0000-0001-9440-8872]{Norman A. Grogin}
\affiliation{Space Telescope Science Institute, 3700 San Martin Dr., Baltimore, MD 21218, USA}

\author[0000-0003-0129-2079]{Santosh Harish}
\affiliation{Laboratory for Multiwavelength Astrophysics, School of Physics and Astronomy, Rochester Institute of Technology, 84 Lomb Memorial Drive, Rochester, NY 14623, USA}

\author[0000-0001-6145-5090]{Nimish P. Hathi}
\affiliation{Space Telescope Science Institute, 3700 San Martin Dr., Baltimore, MD 21218, USA}

\author[0000-0002-4884-6756]{Benne W. Holwerda}
\affil{Physics \& Astronomy Department, University of Louisville, 40292 KY, Louisville, USA}

\author[0000-0001-9298-3523]{Kartheik G. Iyer}
\affiliation{Dunp Institute for Astronomy \& Astrophysics, University of Toronto, Toronto, ON M5S 3H4, Canada}

\author[0000-0001-8152-3943]{Lisa J. Kewley}
\affiliation{Center for Astrophysics, Harvard \& Smithsonian, 60 Garden Street, Cambridge, MA 02138, USA}

\author[0000-0002-5537-8110]{Allison Kirkpatrick}
\affiliation{Department of Physics and Astronomy, University of Kansas, Lawrence, KS 66045, USA}

\author[0000-0002-8360-3880]{Dale D. Kocevski}
\affiliation{Department of Physics and Astronomy, Colby College, Waterville, ME 04901, USA}

\author[0000-0003-2366-8858]{Rebecca L. Larson}
\affiliation{Space Telescope Science Institute, Baltimore, MD 21218, USA}
\affil{Laboratory for Multiwavelength Astrophysics, School of Physics and Astronomy, Rochester Institute of Technology, 84 Lomb Memorial Drive, Rochester, NY 14623, USA}

\author[0000-0003-3130-5643]{Jennifer M. Lotz}
\affiliation{Gemini Observatory/NSF's National Optical-Infrared Astronomy Research Laboratory, 950 N. Cherry Ave., Tucson, AZ 85719, USA}

\author[0000-0003-1581-7825]{Ray A. Lucas}
\affiliation{Space Telescope Science Institute, 3700 San Martin Dr., Baltimore, MD 21218, USA}

\author[0000-0001-9879-7780]{Fabio Pacucci}
\affiliation{Center for Astrophysics, Harvard \& Smithsonian, 60 Garden St, Cambridge, Massachusetts, USA}
\affiliation{Black Hole Initiative, Harvard University, 20 Garden St, Cambridge, Massachusetts, USA}

\author[0000-0001-7503-8482]{Casey Papovich}
\affiliation{Department of Physics and Astronomy, Texas A\&M University, College Station, TX, 77843-4242 USA}
\affiliation{George P.\ and Cynthia Woods Mitchell Institute for Fundamental Physics and Astronomy, Texas A\&M University, College Station, TX, 77843-4242 USA}

\author[0000-0001-8940-6768]{Laura Pentericci}
\affiliation{INAF - Osservatorio Astronomico di Roma, via di Frascati 33, 00078 Monte Porzio Catone, Italy}

\author[0000-0003-4528-5639]{Pablo G. P\'erez-Gonz\'alez}
\affiliation{Centro de Astrobiolog\'{\i}a (CAB), CSIC-INTA, Ctra. de Ajalvir km 4, Torrej\'on de Ardoz, E-28850, Madrid, Spain}

\author[0000-0003-3382-5941]{Nor Pirzkal}
\affiliation{Space Telescope Science Institute, 3700 San Martin Dr., Baltimore, MD 21218, USA}

\author[0000-0002-5269-6527]{Swara Ravindranath}
\affiliation{Astrophysics Science Division, NASA Goddard Space Flight Center, 8800 Greenbelt Road, Greenbelt, MD 20771, USA}
\affiliation{Center for Research and Exploration in Space Science and Technology II, Department of Physics, Catholic University of America, 620 Michigan Ave N.E., Washington DC 20064, USA}

\author[0000-0002-6748-6821]{Rachel S. Somerville}
\affiliation{Center for Computational Astrophysics, Flatiron Institute, 162 5th Avenue, New York, NY, 10010, USA}

\author[0000-0002-1410-0470]{Jonathan R. Trump}
\affiliation{Department of Physics, 196 Auditorium Road, Unit 3046, University of Connecticut, Storrs, CT 06269, USA}

\author[0000-0003-3903-6935]{Stephen M.~Wilkins} %
\affiliation{Astronomy Centre, University of Sussex, Falmer, Brighton BN1 9QH, UK}
\affiliation{Institute of Space Sciences and Astronomy, University of Malta, Msida MSD 2080, Malta}

\author[0000-0001-8835-7722]{Guang Yang}
\affiliation{Nanjing Institute of Astronomical Optics \& Technology, Chinese Academy of Sciences, Nanjing 210042, China}
\affiliation{CAS Key Laboratory of Astronomical Optics \& Technology, Nanjing Institute of Astronomical Optics \& Technology, Nanjing 210042, China}

\author[0000-0003-3466-035X]{{L. Y. Aaron} {Yung}}
\affiliation{Space Telescope Science Institute, 3700 San Martin Drive, Baltimore, MD 21218, USA}



\begin{abstract}
We present the Cosmic Evolution Early Release Science Survey (CEERS) catalog, including space-based photometry, photometric redshifts, and physical parameters for more than 80,000 galaxies. The imaging used for this catalog comes from the CEERS survey, which has NIRCam coverage over $\sim$100 sq.\ arcmin of the Extended Groth Strip (EGS) in seven filters from 1.15\,$\mu$m to 4.44$\,\mu$m. Alongside these data, we also include ancillary HST imaging in seven filters from 0.435\,$\mu$m to 1.6\,$\mu$m. We used Source Extractor with hot and cold detection settings to extract photometry. We derive photometric redshifts using the spectral energy distribution (SED) modeling code, LePHARE, and estimate their accuracy using spectroscopically confirmed galaxies out to $z\sim10$, with $\sigma_{NMAD}$ ranging from 0.035-0.073, depending strongly on galaxy magnitude and redshift. We compute stellar masses, star formation rates, and E(B-V) using three different SED fitting codes with different templates and assumptions about the galaxy star formation histories. All of these measurements, as well as the full mosaics in all filters, and redshift probability distribution functions, are made available via the CEERS DR1.0 data release.


\end{abstract}


\section{Introduction} \label{sec:intro}

In its first three years of observation, JWST \citep{gardner_2006, gardner_2023, rieke23,rigby23} has revolutionized our understanding of the early universe. With its large 6.5\,m mirror and sensitive infrared detectors, it has identified galaxies within the first several hundred million years after the Big Bang (e.g., \citealt{naidu_2022,naidu_2025, castellano2022,finkelstein_2022b, arrabalharo23a,harikane_2023,Curtis-Lake2023,robertson_2023,wang_2023,carniani_2024,witstok_2025}) and found evidence for early supermassive black holes (SMBHs, e.g., \citealt{greene_2023, harikane_2023b, kocevski23b, larson_2023, bogdan_2024, maiolino_2024}). Its ability to resolve detailed structure has also constrained the morphologies of galaxies and our understanding of galaxy growth from Cosmic Dawn to Cosmic Noon (e.g., \citealt{finkelstein_2023,Finkelstein2024,kartaltepe2023,Nelson2023, Ferreira2023,Franco2024,2023ApJ...951L...1P,2025arXiv250315594P}).

One of the first surveys to be observed with JWST was CEERS (PI: S. Finkelstein, PID: \#1345, \citealt{finkelstein2025}), the Cosmic Evolution Early Release Science Survey, a Director’s Discretionary (DD) Early Release Science (ERS) program designed to address some of the outstanding questions in galaxy formation and evolution over a wide range of cosmic history ($0.5<z<12$). The ERS program included 13 approved projects that span all areas of astronomy, with the goal of obtaining observations early in the first year of JWST operations that tested all of the different instrument modes and provided data to the public immediately, thus enabling proposal planning for future cycles.

CEERS utilized four different JWST modes, including NIRCam \citep{rieke23} imaging, MIRI \citep{wright23} imaging, NIRSpec \citep{jakobsen22} MSA spectroscopy, and NIRCam grism spectroscopy, over two observing epochs to validate the efficient use of parallel observations for extragalactic surveys. CEERS targeted the Extended
Groth Strip (EGS) field \citep{davis07}, one of the five deep fields with deep multiwavelength observations from HST through the CANDELS survey \citep{grogin_2011, koekemoer_2011}.

To date, more than 200 publications have utilized CEERS observations, including studies on very high redshift galaxies and the evolution of the UV luminosity function (e.g., \citealt{finkelstein22,finkelstein_2023, Finkelstein2024,arrabalharo23a,fujimoto23,Donnan2023,Endsley2023}), the growth of SMBHs (e.g., \citealt{kocevski23b,kocevski23,larson23a}), and the evolution of galaxy structure and morphology (e.g., \citealt{kartaltepe2023,rose23, rose24, guo23,lebail23,Ono2023,2023ApJ...946L..16P,Robertson2023,Yao2023, pandya24,ward24,Sun2024, guo_2025}).  For a list of all CEERS Key papers, see \citealt{finkelstein2025}. With several Cycle 3 and 4 programs targeting the CEERS field, it is clear that it will have a long-lasting legacy for JWST.

All five of the CANDELS deep fields have now been observed by JWST during Cycle 1, with many follow-up observations in subsequent cycles, cementing their standing as legacy fields. In addition to CEERS observations of EGS, GOODS-N and -S are targeted by JADES \citep{Eisenstein23}, NGDEEP (PIs: S. Finkelstein, C. Papovich, \& N. Pirzkal, PID: \#2079, \citealt{Bagley2024}), and MIDIS (PI: Ostlin, PID 1283 and 6511, \citealt{2025arXiv250315594P}), COSMOS is targeted by COSMOS-Web (PIs: J. S. Kartaltepe \& C. M. Casey, PID: \#1727, \citealt{Casey23}), PRIMER (PI: J. Dunlop, PID: \#1837), and COSMOS-3D (PI: K. Kakiichi, PID: \#5893), and UDS by PRIMER.

In order to facilitate future studies in these legacy fields, precise and accurate photometry, photometric redshifts, and physical parameters are necessary. In this paper, we present the public CEERS NIRCam-selected photometric catalog, which includes photometry from seven NIRCam filters as well as seven existing HST filters, along with photometric redshifts and physical parameters derived from the photometry. We also present the CEERS v1.0 Data Release, with updated NIRCam image reductions, along with the catalog \footnote{\url{https://ceers.github.io/releases.html}}.

\begin{figure*}[!t]
    \centering
    \includegraphics[width=1\linewidth]{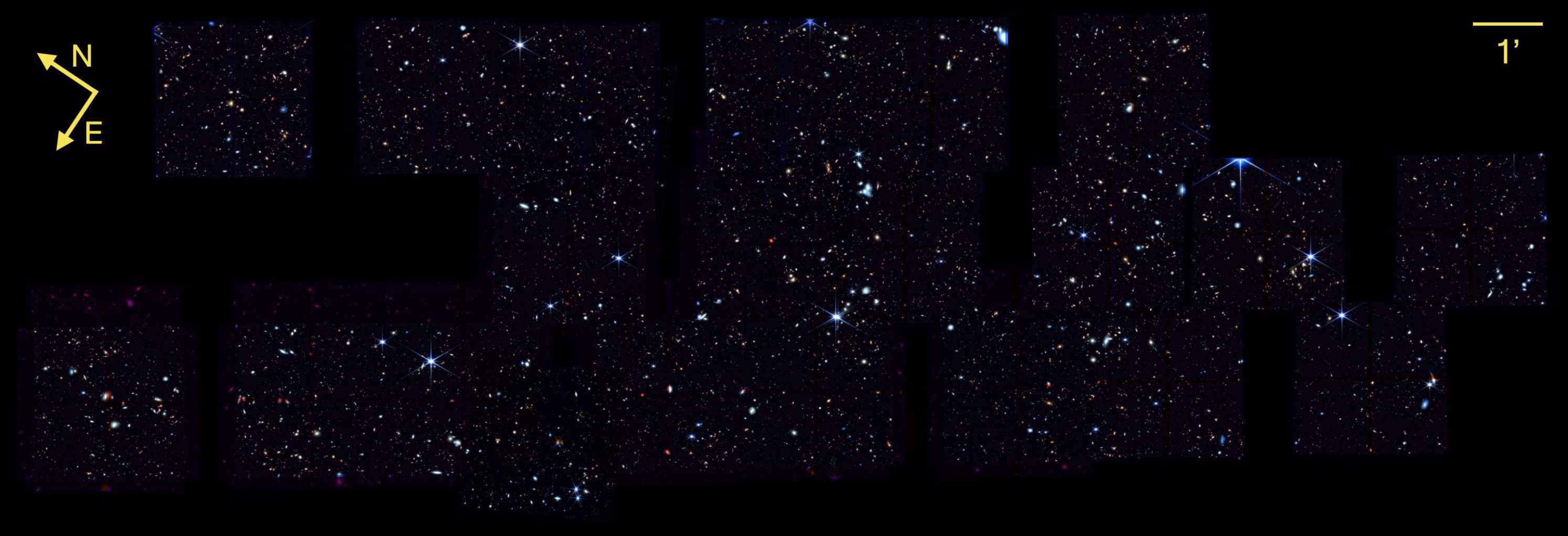}
    \caption{A color composite (RGB combination including all seven NIRCam filters) of the full CEERS v1.0 mosaic, including the half-pointing from DDT program 2750 \citep{arrabalharo23a}. There are regions of the mosaic along the edges of some pointings and in the detector gaps where sources appear redder. These regions are covered by only the long-wavelength NIRCam filters, and we have flagged the sources in this region as described in Section \ref{sec: flags}}
    \label{fig: fullfield}
\end{figure*}

This paper is organized as follows. In Section \ref{sec:data}, we present the JWST and HST observations used. In Section \ref{sec: phot}, we describe the photometric measurements, including aperture and point spread function (PSF) corrections, and empirical error estimation, and validate the photometry in Section \ref{sec: valid}. In Section \ref{sec: redshifts}, we describe the photometric redshift measurements and assess their accuracy, and in Section \ref{sec: phys} we describe other physical parameters. In Section \ref{sec: cat}, we present the catalog and the CEERS v1.0 Data Release. Finally, in Section \ref{sec: summary} we summarize our results. Throughout this paper, all magnitudes are expressed in the AB system \citep{oke_1983}, and we assume the following cosmological parameters: $\mathrm{H_0 = 70 \ km \ s^{-1} \  Mpc^{-1}, \ \Omega_{\Lambda} = 0.7, \ \Omega_{m} = 0.3}$.

\section{Data} \label{sec:data}
\subsection{CEERS NIRCam Imaging}

The CEERS NIRCam imaging comprises ten pointings observed in June and December of 2022, covering a total of $\sim$90 arcmin$^2$, with MIRI or NIRSpec observations as coordinated parallels, designed to maximize the area of overlap between the instruments. The total area and ultimate depth were chosen to achieve the primary science goal of the program--detecting a sizable sample of $z>9$ galaxies in order to constrain the UV luminosity function at these redshifts. Observations were taken in seven filters (F115W, F150W, F200W, F277W, F356W, F410M, F444W) with a total integration time of 2835\,s in each (typically consisting of three dithers with 9 groups in each integration using the MEDIUM8 readout mode), with a second set in F115W, thus doubling the total integration time in this dropout filter. For full details on the observations, see \cite{finkelstein2025}.

We also include the portion of the NIRCam imaging from DDT program 2750 \citep{arrabalharo23a} that overlaps with the other CEERS NIRCam pointings. A single NIRCam pointing with NIRspec prism in parallel was observed in March 2023, and we include one module of the pointing in the CEERS catalog. This program includes the same filters as CEERS, with the exception of F410M. The total integration times are 6345\,s for the F115W and F277W filters and 5701\,s for the remaining filters.

We use the CEERS v1.0 imaging, with the detailed reduction steps described by \citet{bagley2023} from the first epoch of imaging and with updates of the process provided by \citet{finkelstein2025}. Here, we briefly summarize those steps. 

The v1.0 CEERS reduction uses v1.13.4 of the JWST Calibration Pipeline\footnote{\url{https://github.com/spacetelescope/jwst}} \citep{bushouse_2024} with updated CRDS reference files (jwst\_1195.pmap). We use the default pipeline parameters for Stage 1 and the custom snowball removal routine of \cite{bagley2023}. We then perform wisp and 1/f noise\footnote{\url{https://jwst-docs.stsci.edu/known-issues-with-jwst-data/1-f-noise}} removal using the wisp templates of \cite{rieke23b} and \cite{tacchella23b} and the 1/f removal procedure described in Appendix C of \cite{finkelstein2025}. The astrometry was done using the JWST HST Alignment Tool (JHAT, \citealt{jhat}) with an HST/ACS F814W reference catalog tied to Gaia EDR3 \citep{lindegren2021}. We apply custom steps in Stage 3 to improve the handling of outliers, bad pixels, and persistence. The final mosaics are then resampled to a grid with a pixel scale of 0\farcs03/pixel. Finally, we apply a custom script to perform 2D background subtraction on the mosaic in each filter. These final mosaics are used for the catalog described in this paper and including in the v1.0 data release. An RGB color composite image of the v1.0 mosaic is shown in Figure \ref{fig: fullfield}.

\begin{figure*}
    \centering
    \includegraphics[width=1\linewidth]{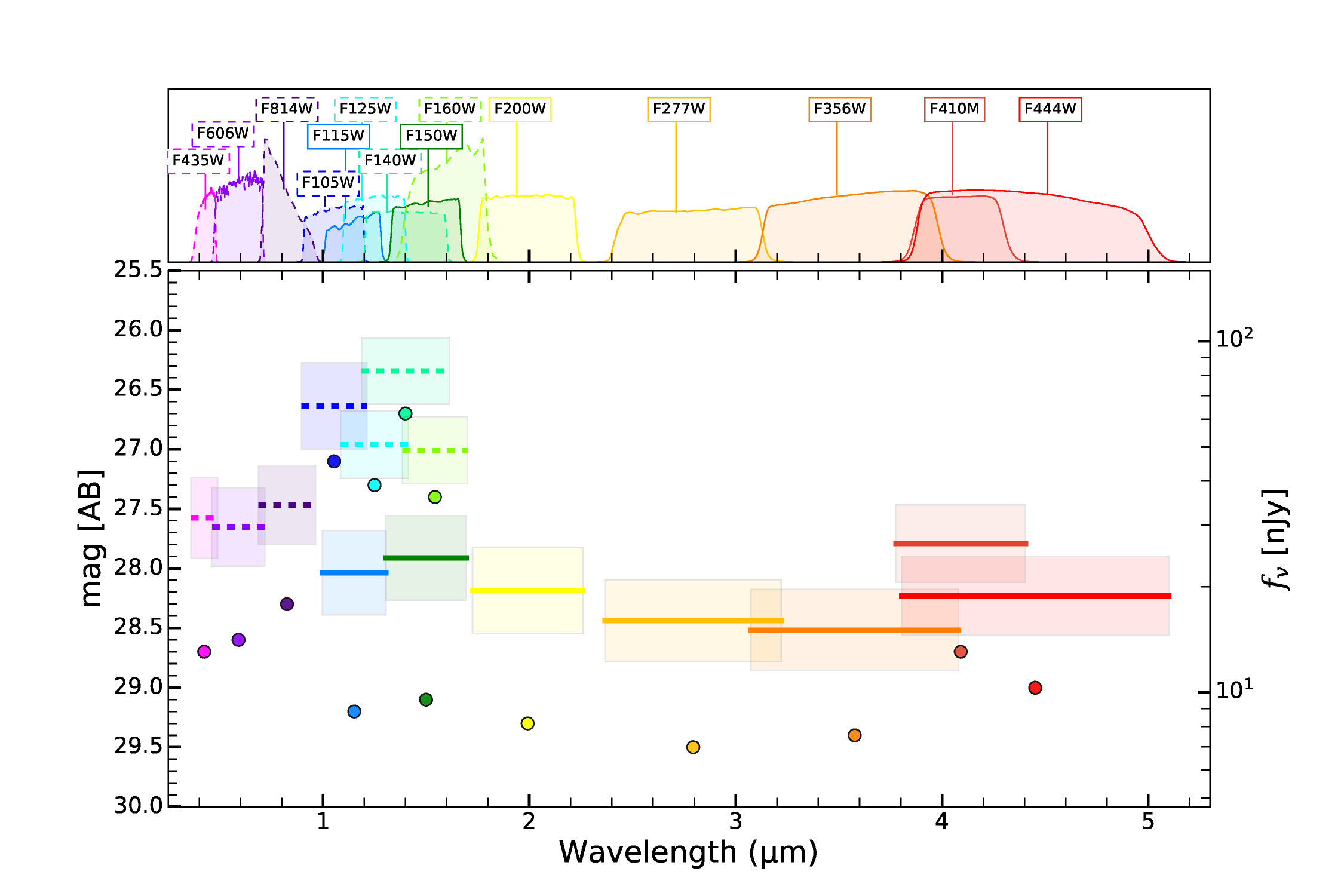}
    \caption{The catalog (lines) and point source (points) depths for each of the 14 photometric filters included in the CEERS photometric catalog along with the filter transmission curves (top panel). Dashed lines represent data for HST filters, whereas solid lines represent JWST filters. The catalog depths were calculated as the 5$\sigma$-clipped median of the empirical errors (described in Section \ref{sec: errors}), and the shaded region shows the 1$\sigma$ spread on each of those measurements. Catalog depths are shallower than point-source depths in each filter due to the resolved nature of most catalog sources. Each of the depth values can be found in Table 4.} 
    \label{fig: filter_depth}
\end{figure*}

\subsection{HST ACS and WFC3 Imaging}

We include the archival HST imaging of the EGS field from the following programs: The All-Wavelength Extended Groth Strip International Survey (AEGIS; \citealt{davis_2007}, ACS F606W and F814W), the Cosmic Assembly
Deep Extragalactic Legacy Survey (CANDELS; \citealt{koekemoer_2011, grogin_2011}, ACS F606W and F814W, WFC3 F125W and F160W), 3D-HST (\citealt{momcheva_2016}, WFC3 F140W), UVCANDELS (\citealt{wang_2025}, ACS F435W), and WFC3 F105W (PID: 13792, PI: R. Bouwens). In total, the CEERS field has complete coverage with ACS F435W, F606W, and F814W, and WFC3 F125W, F140W, and F160W, and partial coverage with WFC3 F105W. To match the CEERS NIRCam mosaics, we use a custom version (v1.9) of the EGS mosaics on the same 0\farcs03/pixel scale as the NIRCam mosaics, and with astrometry aligned to Gaia EDR3. 

Figure \ref{fig: filter_depth} shows the transmission curves of each of the JWST and HST filters included in the catalog, along with the measured catalog depths (see Section \ref{sec: cat_depth}).

\section{Source Detection} \label{sec: phot}

In this section, we describe the process of detecting sources in the CEERS mosaic and measuring their photometry, including the construction of PSFs for each filter, the construction of the NIRCam detection image, the photometric measurements and aperture corrections, and the error estimation.

\subsection{Point Spread Function Matching} \label{sec: psf}

\begin{deluxetable}{cccc}
\tablecaption{The filters included in the catalog, from both HST and JWST, with the FWHM and encircled energy flux for the PSFs produced in each filter, as described in Section \ref{sec: psf}}
\label{table: psf}
\tablehead{\colhead{Telescope / Instrument} & \colhead{Filter} & \colhead{FWHM} & \colhead{PSF Encircled} \\ 
\colhead{} & \colhead{} & \colhead{} & \colhead{Energy (d = 0\farcs2)}} 
\startdata
HST / ACS & F435W & 0\farcs111 & 0.872 \\
HST / ACS & F606W & 0\farcs103 & 0.681 \\
HST / ACS & F814W & 0\farcs109 & 0.607 \\
HST / WFC3 & F105W & 0\farcs225 & 0.332 \\
JWST / NIRCam SW & F115W & 0\farcs050 & 0.743 \\
HST / WFC3 & F125W & 0\farcs229 & 0.313 \\
HST / WFC3 & F140W & 0\farcs233 & 0.301 \\
JWST / NIRCam SW & F150W & 0\farcs060 & 0.745 \\
HST / WFC3 & F160W & 0\farcs238 & 0.285 \\
JWST / NIRCam SW & F200W & 0\farcs075 & 0.707 \\
JWST / NIRCam LW & F277W & 0\farcs121 & 0.601 \\
JWST / NIRCam LW & F356W & 0\farcs139 & 0.553 \\
JWST / NIRCam LW & F410M & 0\farcs154 & 0.517 \\
JWST / NIRCam LW & F444W & 0\farcs160 & 0.494 
\enddata   
\end{deluxetable}

We constructed PSFs for each of the 14 JWST and HST filters using \texttt{PSFEx} \citep{psfex}. We first ran Source Extractor (\texttt{SE}, version 2.28.2; \citealt{bertin_1996}) using each mosaic as both the detection and measurement image, and then selected sources using an initial magnitude cut of $\mathrm{19 \leq m_{AB} \leq 26}$ to avoid including saturated stars.  We use \texttt{PSFEx} to select point sources from this catalog by making magnitude and effective radius cuts to constrain the stellar locus and then construct the PSF for each filter from these point sources. Table \ref{table: psf} shows the full-width at half maximum (FWHM) and encircled energy measurements for the final PSFs generated. In the Appendix, we show these final PSFs for each filter.

The PSF for each NIRCam filter varies in its FWHM. To ensure that a similar fraction of light is measured in each filter, we convolve the mosaics to a uniform PSF. For the NIRCam mosaics, and the three HST ACS mosaics, we PSF-homogenize each filter to F444W (FWHM = 0\farcs160). Since the HST WFC3 PSFs have FWHMs larger than that of F444W, we do not PSF-match them and instead perform a PSF correction as described in Section \ref{sec: psf_corr}.

\begin{figure}
    \centering    
    \includegraphics[width=1\linewidth]{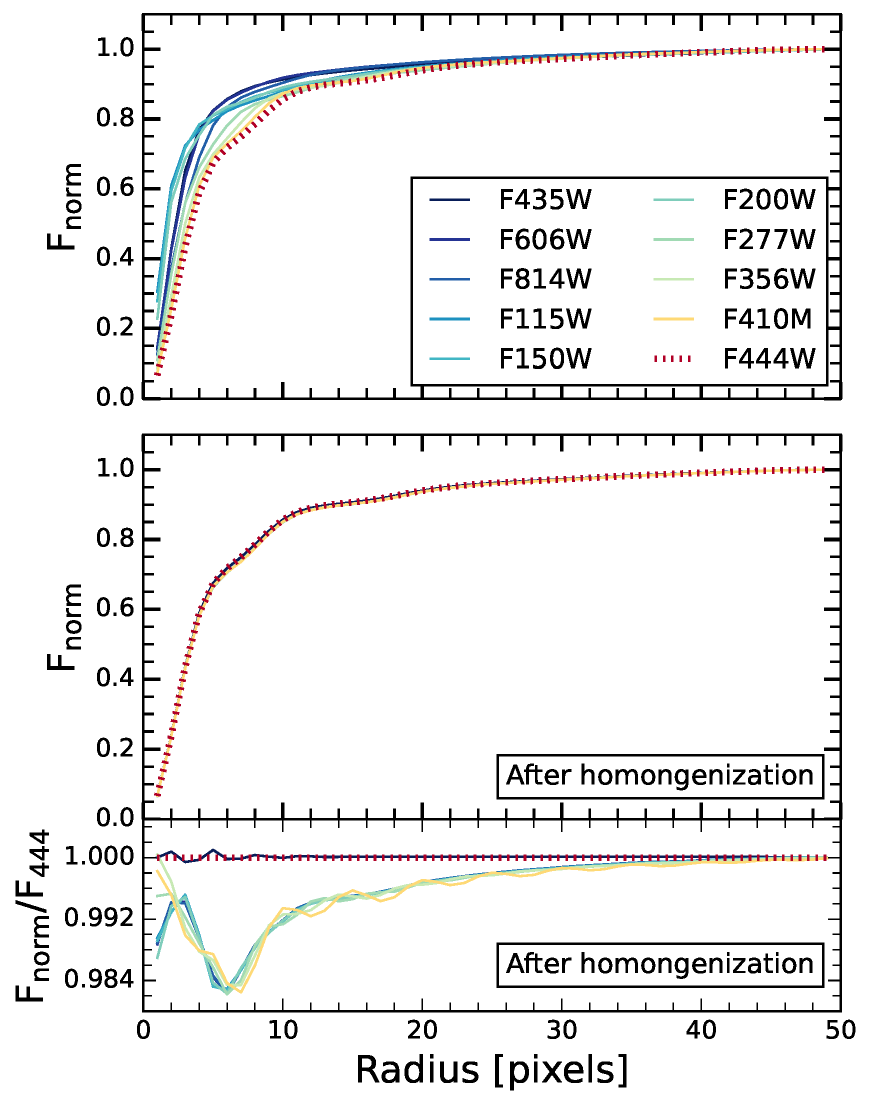}
    \caption{The curves of growth for the PSF models for the nine filters that were PSF-homogenized to F444W (and also including F444W). The top panel shows the normalized fluxes as a function of radius before PSF-homogenization, and the lower panel shows the ratio of the normalized flux to the F444W flux, after homogenization to F444W.}
    \label{fig: psf_convolution}
\end{figure}

The PSF homogenization is performed by generating a convolution kernel using \texttt{PYPHER} \citep{boucaud_2016} and then using the \texttt{convolve} routine in \texttt{IDL} to match the bluer image to the resolution of F444W. Figure \ref{fig: psf_convolution} shows the curves of growth for the 10 PSFs that were homogenized to F444W, and the agreement of their curves of growth after homogenization. All homogenized curves agree within 2\% at all radii.

\subsection{Source Extractor} \label{sec: se}

To construct the photometric catalogs, we run \texttt{SE} in dual-image mode, where we detect sources in a single detection image and make all measurements on the PSF-matched science mosaics. We set the \texttt{WEIGHT\_TYPE} to \texttt{MAP\_RMS} for both the detection image and the science image and use the WHT image output by the JWST pipeline for the science weight (for HST images, we use the RMS maps).

The detection image is the mean $\chi^2$ combination of F150W + F200W + F277W + F356W, which is computed as the square root of the reduced $\chi^2$ \citep{szalay_1999}. The detection image is computed on a per-pixel basis as

\begin{equation}
    \mathrm{I_{detect}(x,y) =\left(\sum{w_i(x,y) s_i^2(x,y)}\right)^{1/2} n(x,y)^{-1/2}},
\end{equation}

\noindent where $w_i$ and $s_i$ are the weight and science images, respectively, for each filter in the combination and $n(x,y)$ is the number of filters covering that pixel. By construction, the corresponding weight map for the mean $\chi^2$ detection image is $\mathds{1}$. We favor using a mean $\chi^2$ detection image over other options, such as an inverse-variance-weighted-sum, because of the non-uniform filter coverage across the CEERS field (in particular due to the long-wavelength observations not always covering the short-wavelength detector gaps, see Section \ref{sec: bad}). We use this detection image for all of our runs of \texttt{SE}. 

For each filter, we perform source detection and photometry using \texttt{SE} in two separate runs, one ``hot” mode and one ``cold" mode to optimize the detection of sources \citep{Rix2004}. This combination allows the detection of fainter, more compact isolated sources (``hot mode") without over-deblending bright extended ones (``cold mode''). This ``hot and cold" catalog strategy is well established and has been used in the construction of previous photometric catalogs (e.g., \citealt{gray_2009, galametz_2013, guo_2013, stefanon_2017, barro_2019, shuntov_2025}).

The hot detection settings were chosen to prioritize the detection of compact sources and faint sources by setting a smaller source minimum area, lower detection threshold, and using a smaller convolution filter. We chose to use a small gaussian kernel for the convolution filter to enhance point-like sources. For the cold mode, we optimized the detection of larger extended sources, while avoiding over-deblending them. For the cold mode, we used a larger (relative to the hot mode) tophat kernel to achieve more uniform smoothing, helping to enhance extended features. The detection settings for both hot and cold modes are given in Table \ref{se_params}. 
\begin{deluxetable}{lccc}
\label{se_params}
\tablecaption{The \texttt{SE} parameter settings as described in Section \ref{sec: se}}
\tablehead{\colhead{Parameter} & \colhead{HOT} & \colhead{COLD}}
\startdata
\texttt{DETECT\_MINAREA} & 5 & 25 \\
\texttt{THRESH\_TYPE} & RELATIVE & RELATIVE \\
\texttt{DETECT\_THRESH} & 6 & 9 \\
\texttt{ANALYSIS\_THRESH} & 6 & 9 \\
\texttt{FILTER\_NAME} & gauss\_3.0\_5x5 & tophat\_4.0\_5x5 \\
\texttt{DEBLEND\_NTHRESH} & 64 & 64 \\
\texttt{DEBLEND\_MINICONT} & 0.00001 & 0.0001 \\
\texttt{CLEAN\_PARAM} & 5.0 & 8.0
\enddata   
\end{deluxetable}

\begin{figure*}[!t]
    \centering
    \includegraphics[trim={2cm 0 0 0},clip,width=1\textwidth]{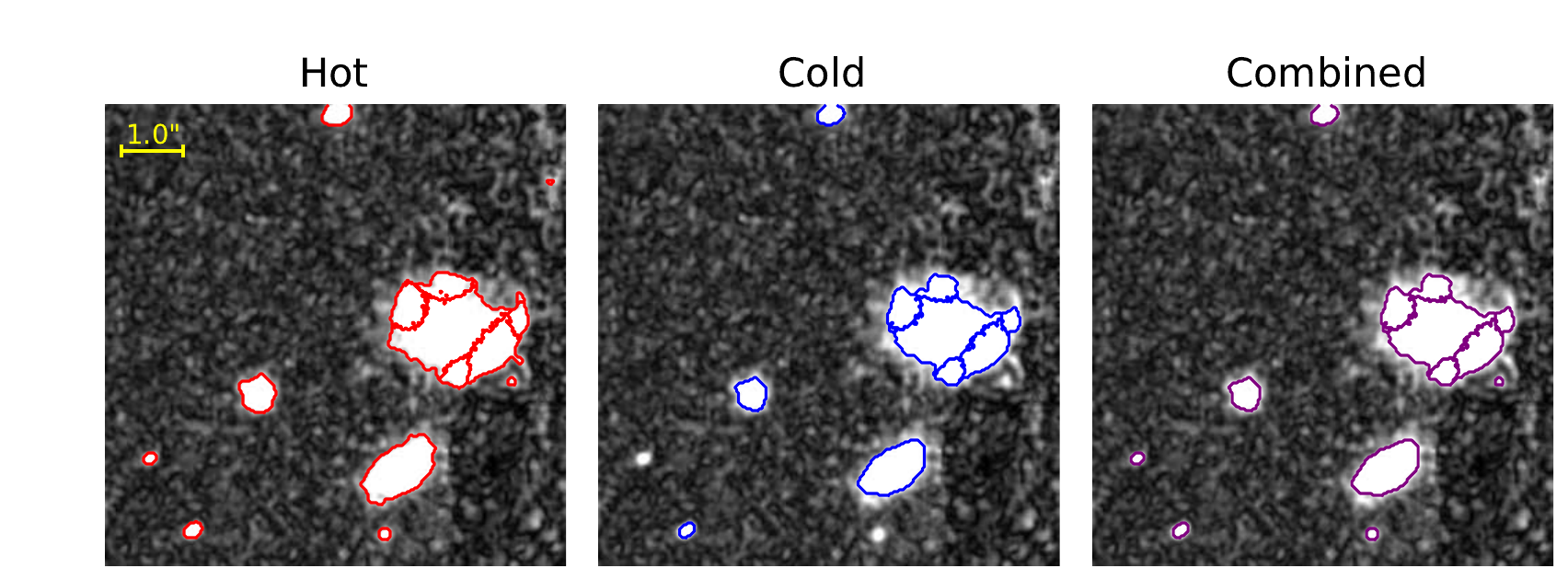}
    \caption{Each panel shows segmentation map (segmap) contours overlaid on the detection image for a small, crowded region of the CEERS field. The left panel shows the segmap contours from the segmap associated with the hot catalog, the middle panel associated with the cold catalog, and the right panel associated with the combined catalog. There are several smaller features in the hot segmap that are not captured in the cold, but are included in the combined based on the combination logic as described in Section \ref{sec: se}. A 1\arcsec\ bar is included for scale. The images are standard North up orientation. We note that the clump in the West part of the image is being appropriately deblended into separate detections and each component is visible as a distinct source with less aggressive scaling.}
    \label{fig: hot_cold}
\end{figure*}
The combination of the hot and cold catalogs follows the logic of the \texttt{IDL} code Galapagos \citep{barden_2012}. In the final catalog, we include all of the cold sources and any hot sources with a centroid point that does not overlap with the Kron radius (dilated by a factor of 5\%) of an existing cold source. This strategy minimizes both the number of spurious sources included in the catalog and the number of extended sources that are inappropriately deblended. We examined the combined segmentation map and determined that the chosen settings are effective in including faint sources while not inappropriately breaking up larger sources. Figure \ref{fig: hot_cold} shows the segmentation map (segmap) bounds over-plotted on the detection image for the hot mode, cold mode, and resultant combined catalog, for a small crowded portion of the field.

As a test for how well the hot and cold strategy is able to detect faint and distant sources, one of CEERS main science goals, we evaluate how well we can recover previously published high-redshift galaxies. Using the 88 candidate z$>$ 8.5 galaxies of \cite{Finkelstein2024}, we successfully recover all of them in the CEERS catalog.

\subsection{Extinction Corrections} \label{sec: extinction}

We then correct the catalog fluxes for Galactic extinction. We assume a value for E(B-V) \citep{schlegel_1998, schlafly_2011} of 0.006 for the EGS field, and compute extinction law values using the \cite{cardelli_1989} Milky Way attenuation curve.

\subsection{Aperture Corrections} \label{sec: aper}

We measure the photometry of sources using a small Kron elliptical aperture in order to ensure accurate colors for higher redshift sources (e.g., \citealt{finkelstein_2010, finkelstein_2022}). \texttt{SE} defines the radius of the aperture with two user-controllable parameters: the ``Kron factor," $k$, and the ``minimum radius," $\mathrm{R_{min}}$.  According to the \texttt{SE} documentation, $k$ is a scale factor that defines how large the aperture will be relative to the Kron radius and $\mathrm{R_{min}}$ is the minimum allowed radius in units of $\mathrm{A\_IMAGE}$. The default values are $k=2.5, \ R_{min} = 3.5$. For our measurements, we use $k = 1.1, \ R_{min} = 1.6$, using the same values as \cite{finkelstein_2022}.

The consequence of using smaller Kron apertures is that the flux measurements are not representative of the total light of a galaxy. We therefore correct the measurements for the flux measured in the Kron apertures by scaling each source's flux by an aperture correction. To compute the aperture correction scale factor, we run our \texttt{SE} setup for the F444W filter, but using the larger default  Kron factor and radius values (2.5, 3.5). The scale factor for each source is computed as the ratio of the F444W-measured flux in the large Kron apertures to the F444W-measured flux in the smaller Kron apertures. This ratio is the multiplicative scale factor we then apply to the original fluxes and flux errors to scale them to a total flux which is then applied to the fluxes in all filters. The median aperture correction factor is 1.54 with a standard deviation of 1.24. 

\subsection{Residual Aperture Correction}
We next perform source injection simulations to account for any remaining flux we are missing due to the wings of the PSF falling below the detection threshold, as noted by previous studies (e.g., \citealt{bouwens_2015, finkelstein_2022}).

We construct images of 3,000 mock galaxies using \texttt{GALFIT} \citep{galfit}. 
The properties of the injected galaxies follow a log-normal distribution of magnitudes from 22 to 30, a log-normal distribution of effective radii centered at $\mu$ = 3.41 pixels with spread of $\sigma$ = 1.8, and a log-normal skewed distribution of S\'ersic indices truncated with $1\leq n \leq 5$. These distributions were chosen to mimic the distributions of real galaxies. We make copies of each of the \texttt{GALFIT} sources convolved separately with the filter PSFs used in the detection image and F444W. The resultant sources convolved with the F444W PSF are then injected into the F444W mosaic for measurements, and these sources for each individual filter are injected into their respective mosaics to create a new detection image with the included mock sources, constructed as described in Section \ref{sec: se}.

We then run \texttt{SE} again on the F444W mosaic with the injected sources (and new corresponding detection image) with the default Kron settings and compare the difference in the recovered magnitudes of these sources with their intrinsic magnitudes. We convert this magnitude difference to a flux density, and fit a linear form of this quantity plotted against the input source magnitude. The fit defines the correction to apply to a source given its magnitude. We calculate a correction factor of 3.5\%  at the bright end ($\mathrm{m_{AB}=22}$), which increases up to 24\% at the faint ($\mathrm{m_{AB}=28}$). We do not extend the fit beyond $\mathrm{m_{AB}=28}$ where the SNR becomes too low to perform a meaningful fit.

\subsection{PSF Corrections} \label{sec: psf_corr}

We do not PSF-match the HST WFC3 mosaics to F444W, because the FWHM's of those PSFs are larger than the FWHM of the F444W PSF. To correct for this, for each of the WFC3 mosaics, we created a new corresponding F444W mosaic PSF-matched to the WFC3 mosaics. We then define a correction factor per source as the native F444W flux divided by the F444W-convolved-to-WFC3 flux. These correction factors are applied multiplicatively to the WFC3 fluxes.

\begin{figure*}[!t]
    \centering
    \includegraphics[width=1\linewidth]{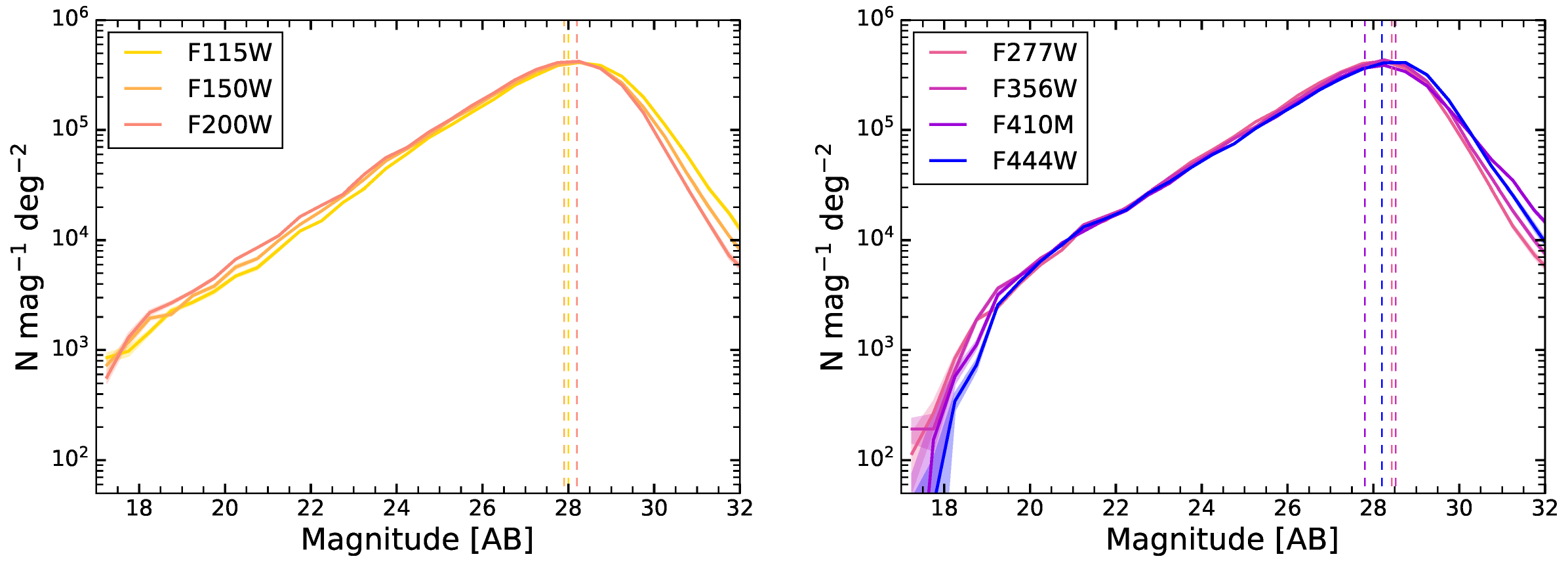}
    \caption{Number counts for the short-wavelength (left panel) and the long-wavelength JWST filters (right panel) included in the CEERS catalog. The shaded regions represent the 1$\sigma$ error from Monte Carlo Bootstrapping. The bin size is 0.5 mag. The vertical dashed lines show the 5$\sigma$ catalog depth for each filter, which each correlate well to the turnover point for each filter, showing that the catalog appropriately includes sources down to 5$\sigma$ depth.}
    \label{fig: numbers}
\end{figure*}

\subsection{Error Estimation} \label{sec: errors}

In addition to the estimation of noise produced by \texttt{SE}, we also opt to calculate the errors empirically since the default errors produced by \texttt{SE} tend to underestimate the true error due to correlated noise. Both sets of errors (from \texttt{SE} and empirical) are kept in the catalog for reference. We calculate the empirical errors following the technique described by \cite{papovich_2016} and \cite{labbe_2003} (see also: \citealt{labbe_2007, whitaker_2011, finkelstein_2022, finkelstein_2023}). We  parameterize a functional form to estimate the noise, $\sigma_n$, of a source that has a linear area $N$,

\begin{equation}
    \sigma_n = \sigma_1 (\alpha N^{\beta} + \gamma N^{\delta}),
    \label{abgd}
\end{equation}

\noindent where $\sigma_1$ is the normalized median absolute deviation $NMAD$ value for the distribution of the sky background, and $\alpha$, $\beta$, $\gamma$, and $\delta$ are the fit parameters. \cite{papovich_2016} let $\delta$ take on any value, enforced that $\alpha$ and $\gamma$ be positive, and $0.5 \leq \beta \leq 1$. With these constraints, the first term of Equation \ref{abgd} represents the case of partially correlated pixels. The case of uncorrelated pixels is $\beta = 0.5$, or $\sigma_n \propto \sqrt{N}$ in the Gaussian limit. Perfectly correlated pixels would be $\beta$ = 1, or $\sigma_n \propto N$.

To measure the noise as a function of area, we compute the $NMAD$, $\sigma_n$, of the distribution of fluxes measured in multiple sizes of non-overlapping, randomly placed circular apertures. We place apertures over the mosaic, avoiding sources (as defined by the segmentation map) and bad pixels in the error maps. We generate 5,000 random, non-overlapping placements to fit 15 apertures of diameters $0\farcs01 \leq d \leq 1\farcs5$ and 500 random, non-overlapping placements to fit 15 apertures of diameter $1\farcs5 \leq d \leq 3\farcs0$. We create a mask to use as a detection image for \texttt{SE} by setting the pixels to one at the center of each aperture, and zero everywhere else. For each aperture size, we measure the noise as the $NMAD$ of the values.

There is a clear dependence of these noise values, $\sigma_n$ on $\sqrt{N}$ as related by a power law. We fit the parameters of Equation \ref{abgd} to our data using an IDL implementation of \texttt{emcee} as described by \cite{finkelstein_2019} and recover parameter values for $\alpha$, $\beta$, $\gamma$, and $\delta$. For all filters, we find $0.5 < \beta < 1$, consistent with partially correlated pixels.

\section{Validation of Photometry} \label{sec: valid}

\subsection{Magnitude Number Counts}

To first assess our photometry, we present the number counts for the JWST filters in the catalog, shown in Figure \ref{fig: numbers}. For this plot, we exclude sources that are flagged as star-contaminated or potentially spurious (see Section \ref{sec: flags}). We compute the number counts by binning sources in bins of 0.5 magnitude width. The counts in each bin represent the number of detected sources per unit area per magnitude, as normalized over the survey areas (0.024 deg$^2$ for the short-wavelength filters and 0.026 deg$^2$ for the long-wavelength filters, with the difference in area being due to the detector gaps in the short-wavelength imaging). In order to estimate the uncertainties on our binned magnitude distribution, we perform Monte Carlo bootstrapping. We find that the magnitude number counts showed the expected trend, with a power-law-like increase from brighter to fainter magnitudes, and then reach a turnover at the point of survey incompleteness.

\subsection{Comparison with the CANDELS EGS Catalog}

\begin{figure*}
    \centering
\includegraphics[width=1\linewidth]{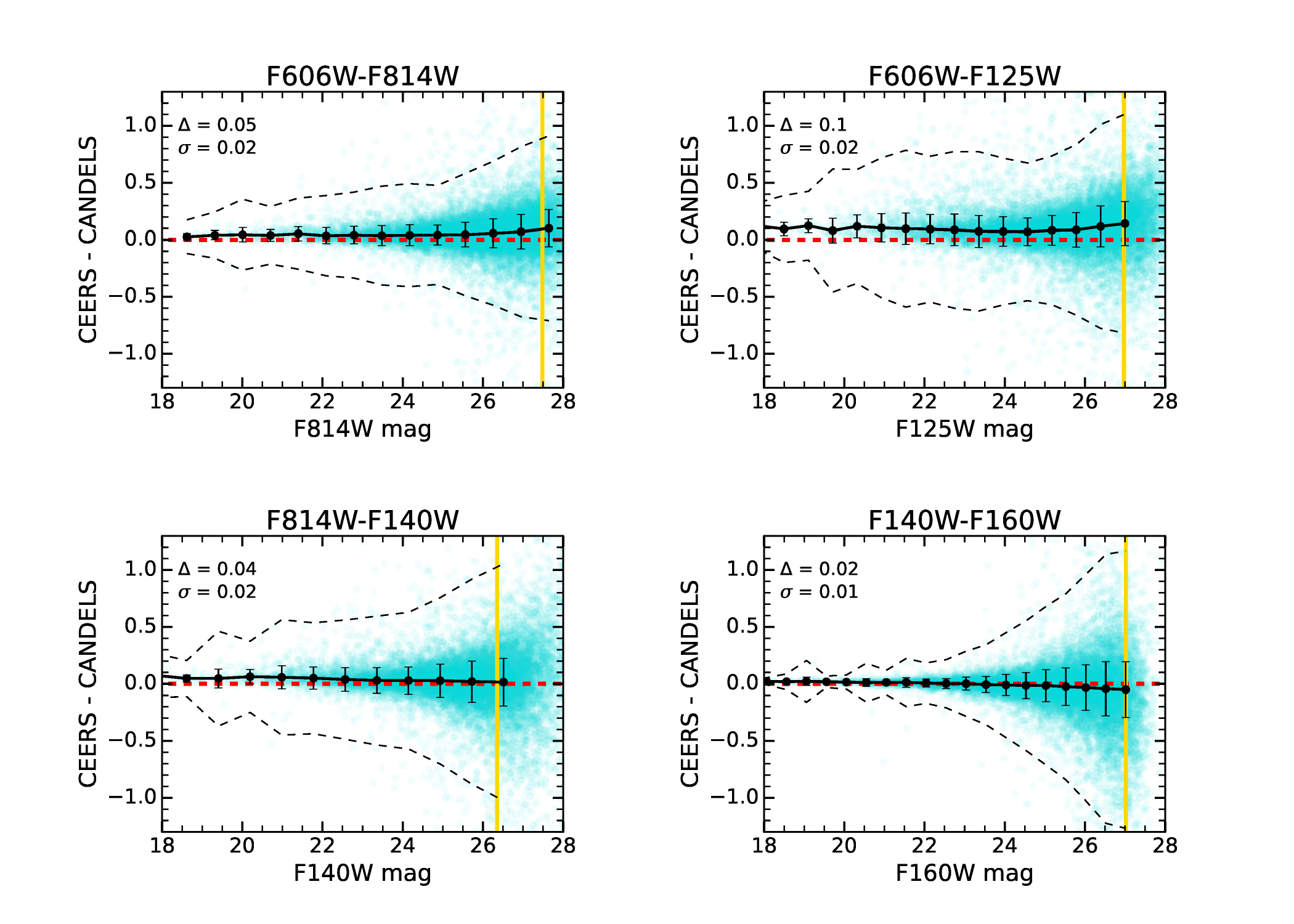}
    \caption{Comparison of colors between the CEERS catalog and the CANDELS catalog for EGS \citep{stefanon_2017}. Individual sources are marked by blue points while the black line and error bars represent the median values and their 1$\sigma$ errors, and the black dashed line shows the 5$\sigma$ envelope. The gold vertical line highlights the 5$\sigma$ CEERS catalog depth for that filter. The $\Delta$ value is the average difference between the line y=0 (red dashed line) and the median line, and the $\sigma$ value is one standard deviation on that. Overall, the agreement between the colors in the CEERS catalog and the CANDELS catalog is good, with a slight systematic offset of 0.1\,mag for F606W-F160W.}
    \label{fig: colors}
\end{figure*}

We compare the CEERS catalog to the CANDELS catalog for EGS, presented by \cite{stefanon_2017}. To assess the consistency of the photometry, we compare colors with the HST filters in common between the CEERS catalog and the CANDELS catalog. We choose to compare colors, rather than the magnitude values, in order to avoid aperture effects. We evaluate F606W-F160W, F814W-F606W, F814W-F140W, and F140W-F160W, shown in Figure \ref{fig: colors}. We choose these specific colors to span the range in wavelength covered by the HST instruments. The comparisons reveal good agreement, with the average difference between the median trend and y=0 line of $<$ 0.1 magnitudes. 

We also evaluate the consistency of the F356W and F444W photometry with the overlapping Spitzer IRAC channels ($\mathrm{3.6\mu m}$ and $\mathrm{4.5\mu m}$). The overall agreement between the JWST photometry and the IRAC photometry is strong, with a difference $<$ 0.1 magnitudes for $m_{AB}$ $<$ 26. At the fainter end, the IRAC photometry is brighter. This is likely caused by over-estimated IRAC fluxes due to blending of multiple sources within the larger PSF. The improved spatial resolution of JWST allows for cleaner separation of objects, splitting up separate neighbors that might have been counted as a single object in the IRAC photometry. Additionally, we note that the upper right panel (F606W - F125W) shows an offset where the CEERS color is 0.1 mag brighter. This is possibly due to the oversampling in our WFC3 images. There is also an offset seen in the lower left panel of F814W-F140W, though not as large, likely because F814W has a slightly larger PSF than F606W.

\begin{figure}
    \centering
    \includegraphics[trim={0 0 0 2cm},clip,width=1\linewidth]{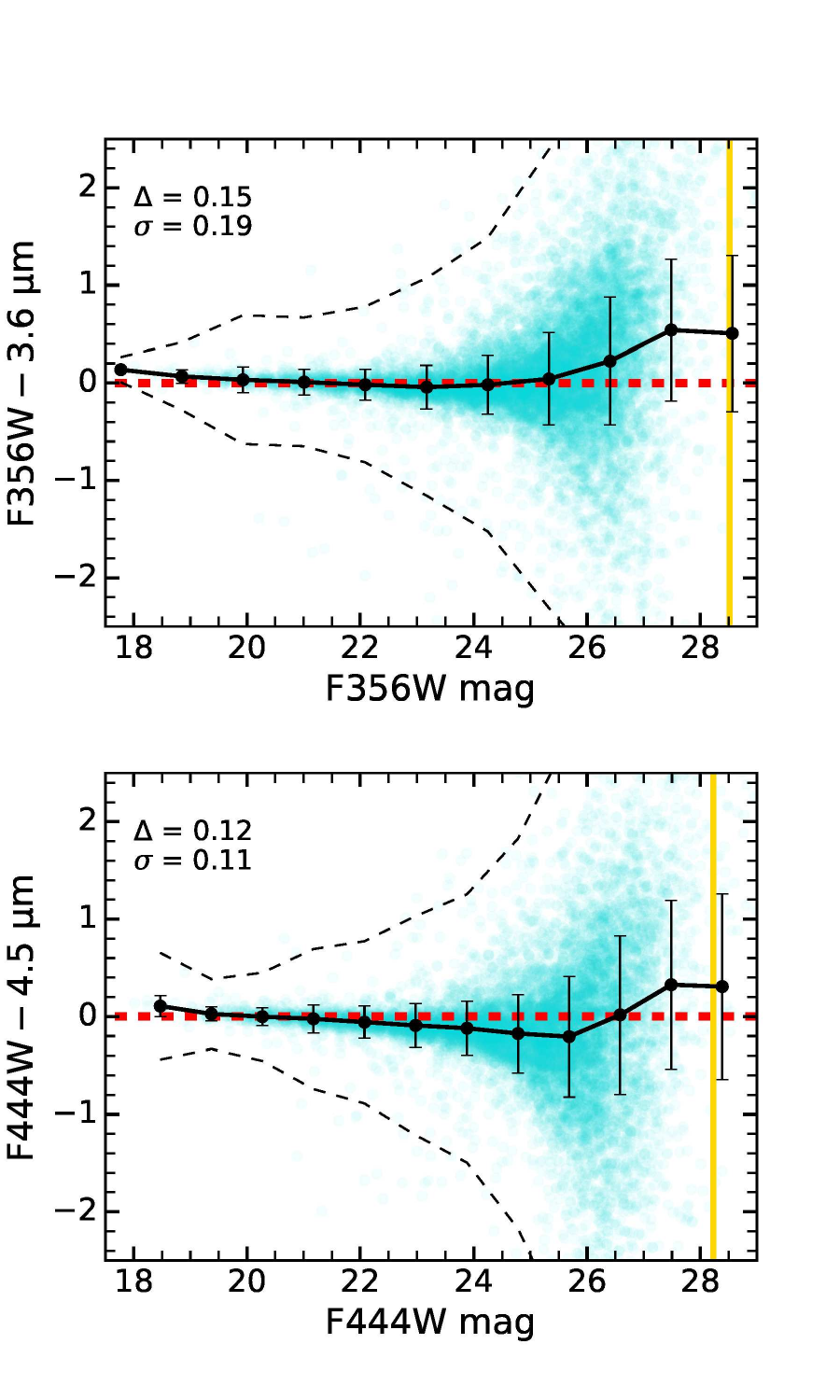}
    \caption{The difference between magnitudes for F356W and 3.6$\mu$m (top panel) and F444W and 4.4$\mu$m (bottom panel) plotted against the respective JWST magnitudes. Individual sources are marked by blue points while the black line and error bars represent the median values and their 1$\sigma$ errors, and the black dashed line shows the 5$\sigma$ envelope. The red dashed line is y=0. The solid gold line highlights the 5$\sigma$ catalog depth for each JWST filter. The 5$\sigma$ depth for the 3.5$\mu m$ and $4.5 \mu m$ are 23.9 and 24.2 respectively (from \citealt{stefanon_2017}). The agreement between the IRAC magnitudes and magnitudes in our CEERS catalog is good, with the average offset $\leq$ 0.09 mags for all filters.}
    \label{fig: irac}
\end{figure}

\section{Photometric Redshifts}
\label{sec: redshifts}

\subsection{Methodology}

We use \texttt{LePHARE} (PHotometric Analysis for Redshift Estimation, \citealt{arnouts1999, ilbert2006}), a template-fitting code for spectral energy distribution (SED) modeling, to estimate the photometric redshifts. \texttt{LePHARE} has been successfully used for photometric redshift estimation for several other large surveys (e.g., \citealt{laigle_2016, weaver_2022, pagul_2024, shuntov_2025}). We include all 14 HST and JWST photometric points in the SED fits. \texttt{LePHARE} uses a \citealt{chabrier_2003} initial mass function (IMF). 

Following \cite{ilbert2015}, we use a library generated with the \cite{bc03} (hereafter; BC03) Stellar Synthesis Population models. The templates included are constructed from six different star formation histories (SFHs) of which four are exponentially declining and the other two are delayed models with a maximum SFR peak occuring after 1 Gyr and 3 Gyr. Two different metallicities are used for each of the six SFHs (solar and half-solar). Additionally, we also run \texttt{LePHARE} with stellar (star) and AGN templates. The stellar templates come from \cite{pickles1998}, \cite{chabrier2000} (which include low-mass stars and brown dwarfs), and \cite{bixler_1991}. The AGN templates are synthetic spectra based on \cite{cristiani_2004}.

We prioritize selecting a diverse set of parameter values to apply to the models in order to span the full range of galaxy populations. For dust attenuation, we apply three different attenuation curves to models \citep{calzetti2000, arnouts2013, salim2018} and 13 E(B-V) values with a maximum value of 1.2. We restrict the parameter space to 43 ages, spanning 0.01 - 13.5 Gyr (spaced logarithmically), in order to accommodate a reasonable computational run time, and 1,500 redshift steps. Our parameter space is made up of all of the combination of these property values, and includes more than 30 million models. We add emission lines to the models as described by \cite{saito_2020}, allowing the emission line flux to vary by a factor up to $\pm$0.5, 1, or 2 of the standard flux computed by the model \texttt{LePHARE} uses from \cite{schaerer_2009}. 

As an initial step before running \texttt{LePHARE} on the full sample of galaxies, we run \texttt{LePHARE} on a subsample of 1,525 sources with fixed spectroscopic redshifts and \texttt{AUTO\_ADAPT} on. This function follows the procedure of \cite{ilbert2006} to adapt the auto calibration of the photometry. The systemic offsets are then calculated between the observed photometry and model photometry at the sources' redshifts and corrections are applied iteratively to the photometric zeropoints to minimize the offsets. We find that the correction for all of the filters is less than 0.05 mag and the final photometric redshift values change minimally with this adaption.
    
\subsection{Photometric Redshift Assessment}

We assess the performance of the photometric redshift fits by comparing them to a compiled list of high-quality flag spectroscopic redshifts. 1,553 of those spectroscopic redshifts are from a compilation of ground-based and HST grism measurements and 1,359 are from CEERS \citep{arrabalharo23a} and RUBIES \citep{deGraaff_2025}. Quantitatively, we measure the performance of the photometric redshift fits by using the $NMAD$ and the outlier fraction ($\eta$). We calculate the $NMAD$ as 
\begin{equation}
    \mathrm{1.48 \times median \left( \frac{\Delta z - median (\Delta z)}{1 + z_{spec}}\right)},
\end{equation}

\noindent where $\Delta z$ is defined as the difference $\mathrm{z_{phot} - z_{spec}}$. For the fraction of outlier sources, $\eta$, we consider a source to be an outlier when $\left| \Delta z \right| > 0.15 \,(1 + z_{\mathrm{spec}})$
 as defined by \cite{hildebrandt2012}. 

\begin{figure*}[!t]
    \centering
\includegraphics[trim={2.5cm 0 2.5cm 0},clip,scale=0.55]{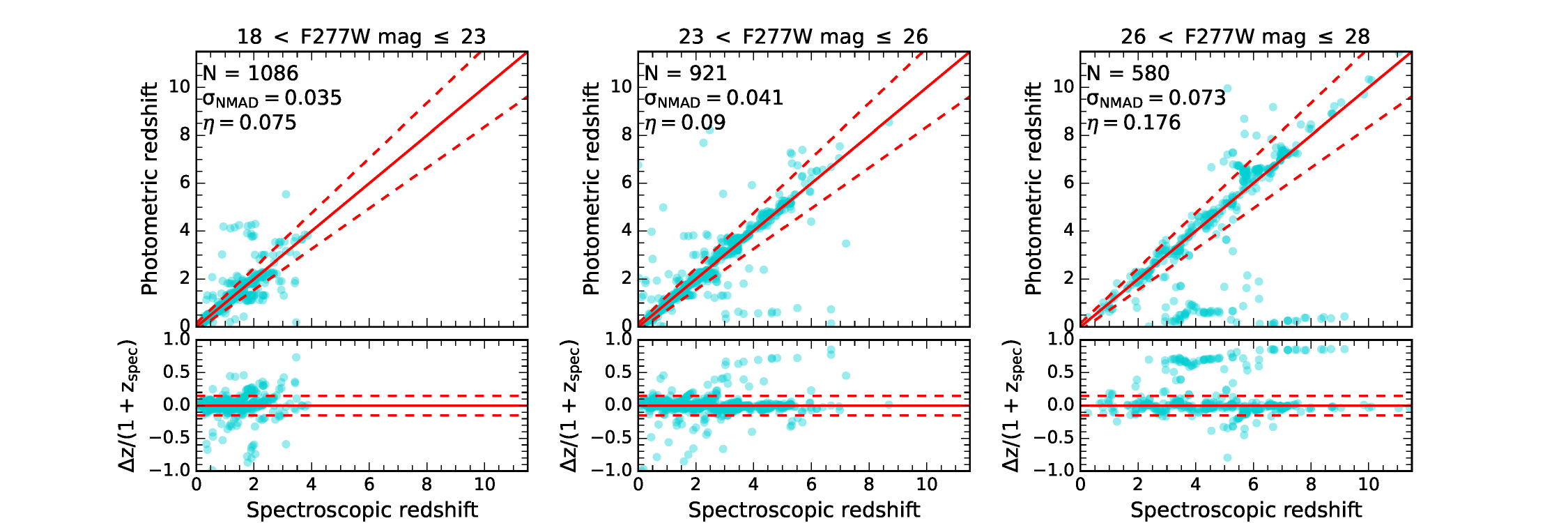}
        \caption{The agreement between the \texttt{LePHARE} photometric redshifts and existing spectroscopic redshifts for the CEERS coverage of the EGS field. The solid red line is the line of equality and the dashed red lines identify where $\left| \tfrac{\Delta z}{1+z_{\mathrm{spec}}} \right| = \pm 0.15$, demarcating where sources are considered outliers. Sources are divided into panels based on their F277W magnitudes. In total, 1,525 sources in the catalog matched to sources in the spectroscopic catalog. Overall, there is good agreement between the redshifts, with an increase in the $NMAD$ and $\eta$ values with increasing magnitude, as expected.}
    \label{fig: redshifts}
\end{figure*}

Figure \ref{fig: redshifts} shows the agreement between the photometric redshifts and the existing spectroscopic redshifts in three different F277W magnitude bins for the CEERS field. Overall, the agreement is generally good, with $NMAD$ values of 0.035 at 18$<$F277W$<$23, 0.041 for 23$<$F277W$<$26, and 0.073 at $26>$F277W$<28$ and $\eta$ values of 0.075, 0.09, and 0.176. The median magnitude value in each bin (18$<$F277W$<$23, 23$<$F277W$<$26, $26>$F277W$<28$) is 21.82, 23.89, and 27.05, respectively. The increasing outlier fraction with source magnitude is expected. For the faintest magnitude bin, we see a significant number of sources with photometric redshifts that are much lower than their predicted spectroscopic redshifts. Many of these sources have a second peak in the photometric redshift PDF that agrees closely with the spectroscopic redshift. We therefore caution users that they should take the full PDF into account, particularly for faint sources. For both the redshifts and the \texttt{LePHARE} physical parameter values (described below), we report the \texttt{MEDIAN} value (of the PDF$_z$) and the \texttt{BEST} value, where \texttt{BEST} is the minimum $\chi^2$ value of the PDF$_z$.

\section{Physical Parameters} \label{sec: phys}
We provide physical parameter measurements, including stellar mass (M$_\star$), star formation rate (SFR), and dust content (E(B-V)) using three different SED fitting codes, as described below. Since each code uses different assumptions for the SFH and different template library, we opt to make all three sets of measurements available so that the user can select the one that best matches their scientific needs. For all three sets of measurements, we fix the redshift to the median of the photometric redshift probability distribution function (PDF$_z$) from \texttt{LePHARE} for consistency. 

\subsection{LePHARE Methodology} \label{sec: phys_method}

\begin{figure}
    \centering
    \includegraphics[width=1.1\linewidth]{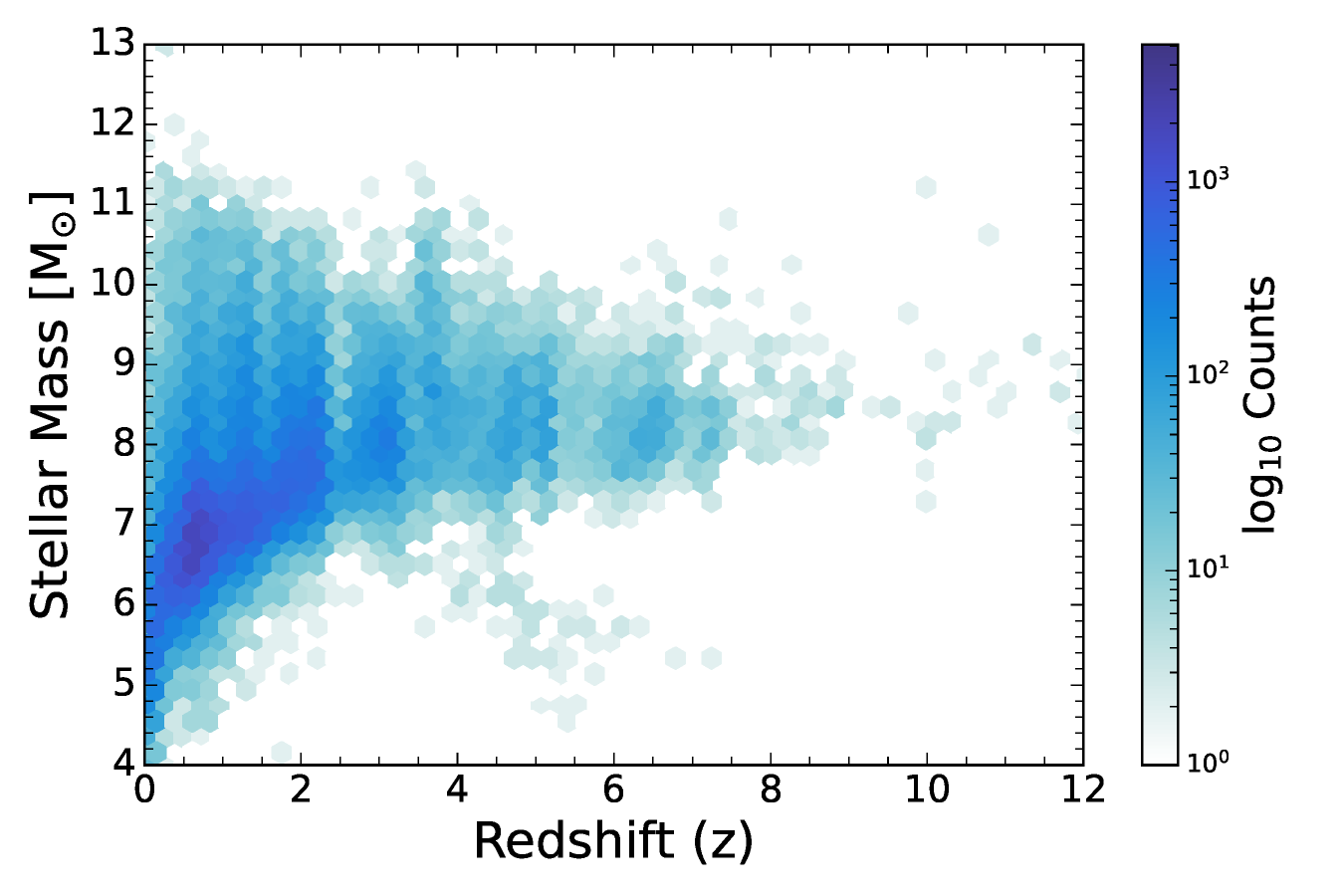}
        \caption{The median stellar masses derived for galaxies in the CEERS catalog using \texttt{LePHARE} as a function of their photometric redshifts.}
    \label{fig: z_mstar}
\end{figure}

We use the same SED template library described above for the photometric redshift measurements to measure physical parameters with \texttt{LePHARE}. To reiterate, this library includes a combination of SFHs, including exponentially declining and delayed models. We compute errors on each these parameters using the PDFs for each parameter and note that these errors do not propagate uncertainties on the photometric redshifts. Figure \ref{fig: z_mstar} shows the distribution of stellar mass versus redshift for our sample, showing an overall smooth distribution of sources. We note the large clump of sources at $z\sim12-14$ and warn the user that these sources have not been individually vetted and may contain spurious sources or have erroneous photometric redshifts.

\subsection{CIGALE Methodology} \label{sec: cigale}

We also use \texttt{CIGALE} \citep{burgarella_2005, noll_2009, boquien_2019} to derive physical parameters. CIGALE is a Python code that generates a grid of models that are compared to each SED and estimates physical properties using the resultant likelihood distribution. Here, we briefly describe the key assumptions used for our \texttt{CIGALE} run. We assume a \cite{chabrier_2003} IMF with BC03 stellar population synthesis models and the following values of stellar metallicities: 0.0004, 0.004, 0.008, 0.02, 0.05. For the SFH, we assume a delayed-$\tau$ model with two components. The first component is the majority old stellar population and the second component is a burst appended on top with $\tau_{burst}$ between 5 to 100 Myr.

\subsection{Dense Basis Methodology} \label{sec: db}

We also  run \texttt{Dense Basis} \citep{Iyer_2017,Iyer_2019} to estimate physical parameters. To reconstruct SFHs, \texttt{Dense Basis} implements a flexible non-parametric SFH represented by a Gaussian Mixture Model \citep[GMM; ][]{Iyer_2019}. The SFHs are built using stellar templates generated from \texttt{FSPS} \citep{Conroy2009, ConroyGunn2010} including implementation of nebular emission lines using \texttt{CLOUDY} \citep{Ferland2017, Byler2017}. For this work, three ``shape'' parameters were used to describe the SFH for the GMM: $t_{25}, t_{50},$ and $t_{75}$ (requiring the recovered SFH of the galaxy to form ``$x$'' fraction of its total mass by time $t_x$). We assume a  \cite{chabrier_2003} IMF and \citet{calzetti2000} dust law. We impose a uniform (flat) prior on the specific SFR (sSFR) with limits on the sSFR $\textrm{yr}^{-1} \in [-14, -7]$, an exponential prior on the dust attenuation over a wide range of values ($A_V \in [0, 4]$), and a uniform (in log-space) prior on the metallicity ($Z/Z_\odot \in [0.01, 2.0]$). We incorporate the photometric redshift from \texttt{LePHARE} as a prior by restricting the redshift space to a top-hat function within the 68 \% confidence interval.

\subsection{Stellar Mass Comparisons} \label{sec: phys_asssess}

In order to assess the reliability of the derived physical parameters, we conduct comparisons among each available stellar mass measurement. We first compare the \texttt{LePHARE} stellar masses to the stellar masses included in the \citep{stefanon_2017} catalog for the CANDELS EGS field, and then to those derived using (\texttt{CIGALE} and \texttt{Dense Basis}). 

\begin{figure*}[!t]
    \centering
    \includegraphics[width=0.32\textwidth]{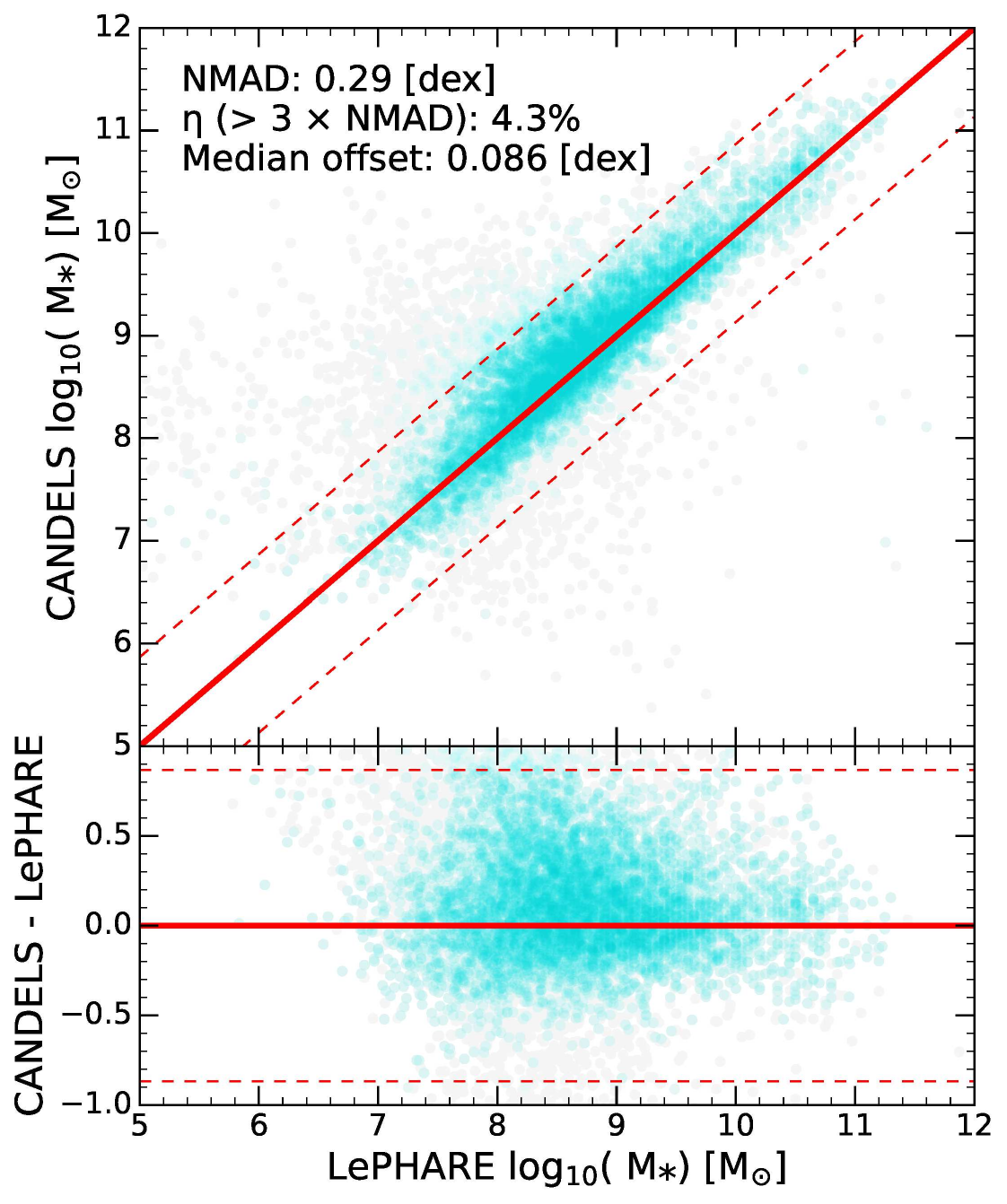}\hfill
    \includegraphics[width=0.32\textwidth]{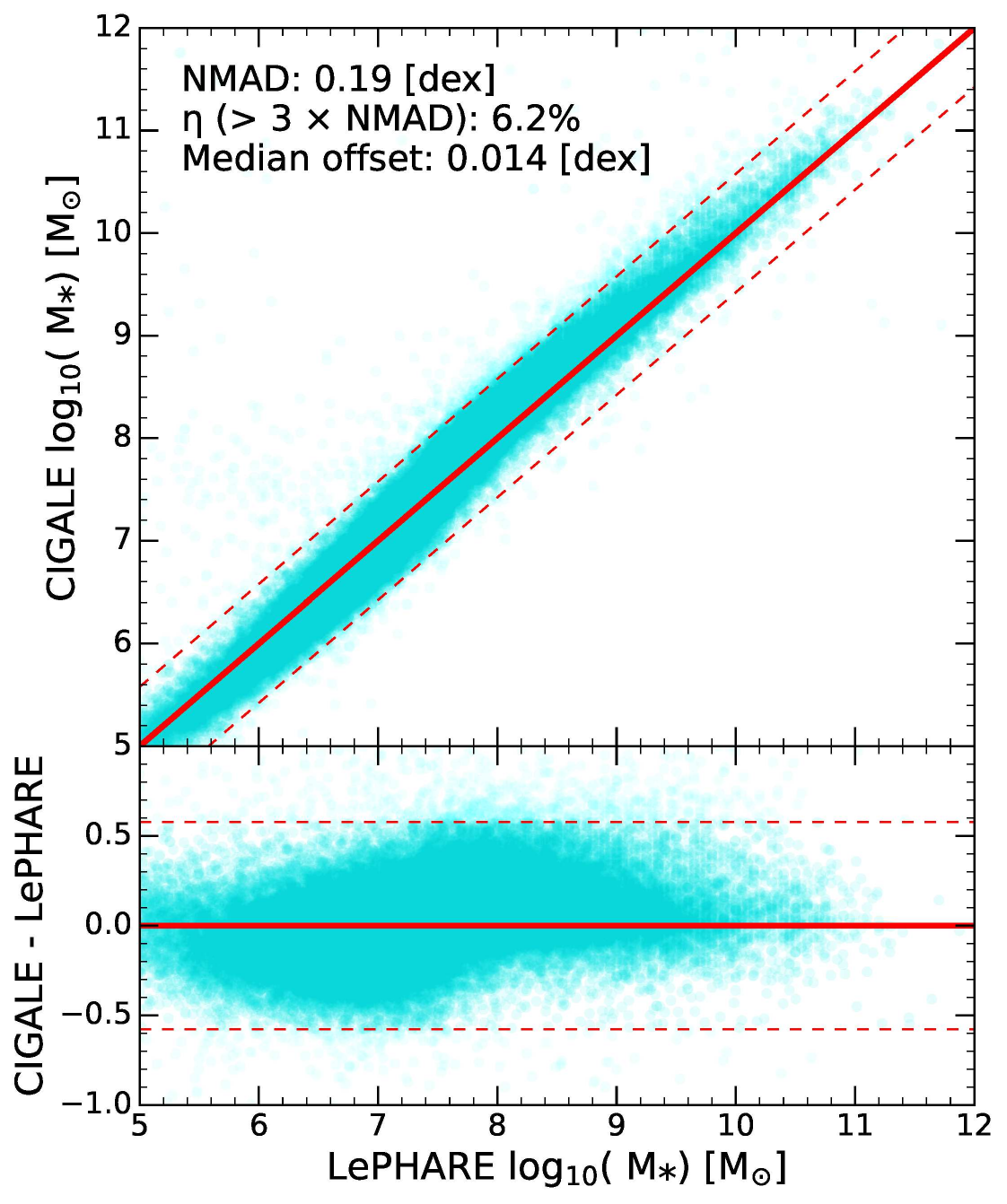}\hfill
    \includegraphics[width=0.32\textwidth]{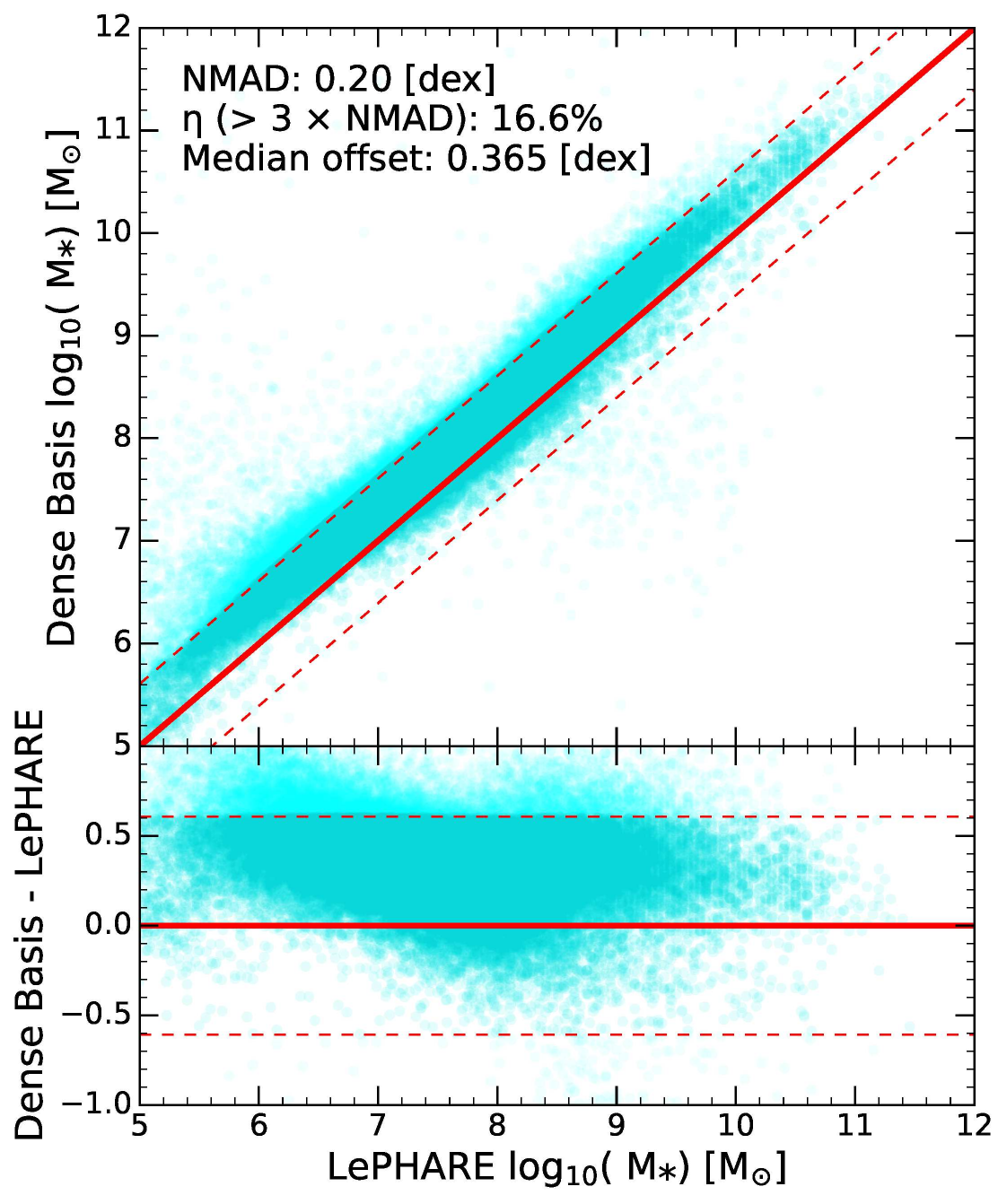}

    \caption{CANDELS median stellar mass from \cite{stefanon_2017} (left),  \texttt{CIGALE} stellar mass (center), and \texttt{Dense Basis} stellar mass (right) versus \texttt{LePHARE} stellar mass. Each panel includes a solid red line that is the line of equality, dashed red lines demarcate outliers (computed as 3$\times NMAD$), and statistics in the upper left show the $NMAD$, outlier fraction, and median offset (excluding outliers). For the CANDELS comparison, galaxies with photometric redshift $> 0.1\times (1+z)$ are shown in gray and excluded from the statistics.}
\label{fig:cat_mass_comp}
\end{figure*}

The CANDELS catalog includes multiple estimates of stellar mass, based on different SED modeling setups (using different codes and SFH assumptions). We use the computed median stellar mass of all of the measurements for each source for our comparison. The left panel of Figure \ref{fig:cat_mass_comp} shows the comparison between the CANDELS stellar masses and the \texttt{LePHARE} stellar masses, for all 10,572 sources that match between the two catalogs, while for the statistics, we include only the 6,495 sources (highlighted in cyan) with photometric redshifts with $\frac{|\Delta_z|}{1+(z_1+z_2)/2} < 0.1$, where $z_1$ is the \texttt{LePHARE} redshift and $z_2$ is the CANDELS redshift. We include only these sources with close photometric redshifts to avoid biasing the mass comparison with sources that had disparate redshifts.  The comparison shows that there is reasonable agreement between the CANDELS median stellar masses and the \texttt{LePHARE} CEERS stellar masses, with an $NMAD$ value of 0.29 dex, and a median offset 0.086 dex (excluding catastrophic outliers, defined as $> 3\times$ $NMAD$), and an outlier fraction of 4.3\%.

We also compare the \texttt{LePHARE} stellar masses to those measured with \texttt{CIGALE}, as described in Section \ref{sec: cigale}. The two measurements are reasonably consistent with one another, with an $NMAD$ value of 0.18 dex, and a median offset 0.009 dex, and an outlier fraction of 2\%. We note that at low stellar masses ($\lesssim 8.5\,M_{\odot}$) the \texttt{LePHARE} masses are slightly smaller and at higher stellar masses, slightly larger. This effect can be explained by the different treatments of nebular emission line modeling between the two codes. For low stellar masses, \texttt{CIGALE} models stronger emission lines, pushing continuum levels lower and thus yielding lower stellar masses, with the reverse happening at higher stellar masses. We note however, that this effect is small. 

Finally, we compare the stellar masses from \texttt{Dense Basis} to those from \texttt{LePHARE}. This comparison has a similar level of scatter as the \texttt{CIGALE} comparison, with an $NMAD$ value of 0.19 dex. However, it has a significant offset of 0.358 dex, with \texttt{Dense Basis} stellar masses being categorically larger than \texttt{LePHARE} stellar masses, and an outlier fraction of 16.7\%. We attribute this difference to \texttt{Dense Basis} using a non-parametric SFH, which gives more weight to older stellar populations and thus increases the measured stellar masses. A similar trend is seen in the literature when comparing stellar masses using parametric and non-parametric SFHs (e.g., \citealt{carnall_2019, leja_2019, lower_2020, whitler_2023}).

\section{The CEERS Catalog} \label{sec: cat}
Here, we present and release the final CEERS source catalog. Table \ref{tab:columns} describes the columns that are included in the catalog, which comprehensively includes photometry, photometric redshifts, and physical parameters for 87,246 sources. We note that previous work published a photometric catalog in CEERS based on an earlier version of the CEERS imaging (v0.51) as part of the ASTRODEEP project \citep{merlin_2024}. The detection strategy of the ASTRODEEP catalog was optimized to detect faint, high-redshift galaxies, with a weighted stack of F356W and F444W as the detection image (and excluding short wavelength filters) and a single set of SE parameters, rather than the hot+cold strategy adopted here. We also note that while the ASTRODEEP catalog includes photometric redshifts, it does not include other physical parameters, such as stellar mass and SFR.
\startlongtable
\begin{deluxetable*}{lll}
\tablecaption{Photometric Catalog Column Descriptions\label{tab:columns}}
\tablehead{
\colhead{Column Name} & \colhead{Description} & \colhead{Units}
}
\startdata
NUMBER & Identification number from Source Extractor& \\
X\_IMAGE & $x$-coordinate of the source centroid (detection image)& Pixels \\
Y\_IMAGE & $y$-coordinate of the source centroid (detection image)& Pixels \\
RA & J2000 right ascension of centroid (detection image)& \\
DEC & J2000 declination of centroid (detection image) & \\
CXX\_IMAGE & Elliptical parameter CXX from SE & Inverse square pixels\\
CXY\_IMAGE & Elliptical parameter CXY from SE & Inverse square pixels\\
CYY\_IMAGE & Elliptical parameter CYY from SE & Inverse square pixels \\
THETA\_IMAGE & Position angle of the Kron aperture &  Degrees \\
ELLIPTICITY & 1 - B\_IMAGE / A\_IMAGE&\\
KRON\_RADIUS & Kron radius for small fiducial aperture& \\
A\_IMAGE & Semi-major axis of Kron aperture is: A\_IMAGE $\times$ KRON\_RADIUS & Pixels \\
B\_IMAGE & Semi-minor axis of Kron aperture is: B\_IMAGE $\times$ KRON\_RADIUS & Pixels \\
ISO\_AREA & Isophotal area from segmentation map & Pixels\\
STAR\_FLAG & Flag indicating source overlaps star mask & \\
BAD\_REGION\_FLAG & Flag indicating source lies in potentially spurious region & \\
APCORR\_AUTO & Aperture correction factor applied to AUTO fluxes & \\
APCORR\_RESIDUAL & Residual aperture correction for AUTO and circular apertures & \\
F\#\#\#\_FLUX & Kron flux (with aperture correction) in filter \#\#\#& nJy\\
F\#\#\#\_FLUXERR\_SE & SE-reported error on Kron flux (with corrections) & nJy\\
F\#\#\#\_FLUXERR\_EMP & Empirical error on Kron flux (with corrections) & nJy\\
F\#\#\#\_MAG & Kron mag (with aperture correction) in filter \#\#\#& mag\\
F\#\#\#\_MAGERR\_SE & SE-reported error on Kron mag (with corrections)& mag\\
F\#\#\#\_MAGERR\_EMP & Empirical error on Kron mag (with corrections)& mag\\
F\#\#\#\_FLUX\_APER & Flux in 12 circular apertures (with residual correction) & nJy\\
F\#\#\#\_FLUXERR\_APER\_SE & SE error in 12 circular apertures (with residual correction) & nJy\\
F\#\#\#\_FLUXERR\_APER\_EMP & Empirical error in 12 circular apertures (with residual correction) & nJy\\
F\#\#\#\_CLASS\_STAR & CLASS\_STAR value from SE (likelihood of stellar source)& \\
LP\_Z\_MED & Median photometric redshift (\texttt{LePHARE} PDF)& \\
LP\_Z\_MED\_LOW & $-1\sigma$ error on $z_\mathrm{phot}$ (16th percentile from \texttt{LePHARE} PDF)& \\
LP\_Z\_MED\_HIGH & $+1\sigma$ error on $z_\mathrm{phot}$ (84th percentile from \texttt{LePHARE} PDF)& \\
LP\_Z\_BEST & $\chi^2$ minimized redshift (\texttt{LePHARE} PDF)& \\
LP\_Z\_CHI & $\chi^2$ value of best-fit \texttt{LePHARE} model& \\
LP\_NBANDS & Number of bands used for  \texttt{LePHARE} fit& \\
LP\_MASS\_MED & \texttt{LePHARE} Median stellar mass & $\mathrm{M_{\odot}}$\\
LP\_MASS\_MED\_LOW & \texttt{LePHARE} $-1\sigma$ error on mass (16th percentile)& $\mathrm{M_{\odot}}$\\
LP\_MASS\_MED\_HIGH & \texttt{LePHARE} $+1\sigma$ error on mass (84th percentile)& $\mathrm{M_{\odot}}$\\
LP\_MASS\_BEST & \texttt{LePHARE} $\chi^2$ minimized stellar mass & $\mathrm{M_{\odot}}$\\
LP\_SFR\_MED & \texttt{LePHARE} Median star formation rate& $\mathrm{M_{\odot} \ yr^{-1}}$\\
LP\_SFR\_MED\_LOW & \texttt{LePHARE} $-1\sigma$ error on star formation rate (16th percentile)& $\mathrm{M_{\odot} \ yr^{-1}}$\\
LP\_SFR\_MED\_HIGH & \texttt{LePHARE} $+1\sigma$ error on star formation rate (84th percentile)& $\mathrm{M_{\odot} \ yr^{-1}}$\\
LP\_SFR\_BEST & \texttt{LePHARE} $\chi^2$ minimized star formation rate & $\mathrm{M_{\odot} \ yr^{-1}}$\\
LP\_E\_BV & \texttt{LePHARE} E(B-V) value from best-fit template & mag \\
DB\_MASS & \texttt{Dense Basis} 50th percentile Bayesian stellar mass  & $\mathrm{M_{\odot}}$\\
DB\_MASS\_LOW & \texttt{Dense Basis} 16th percentile Bayesian stellar mass  & $\mathrm{M_{\odot}}$\\
DB\_MASS\_HIGH & \texttt{Dense Basis} 84th percentile Bayesian stellar mass  & $\mathrm{M_{\odot}}$\\
DB\_SFR & \texttt{Dense Basis} 50th percentile Bayesian SFR  & $\mathrm{M_{\odot} \ yr^{-1}}$\\
DB\_SFR\_LOW & \texttt{Dense Basis} 16th percentile Bayesian SFR & $\mathrm{M_{\odot} \ yr^{-1}}$\\
DB\_SFR\_HIGH & \texttt{Dense Basis} 84th percentile Bayesian SFR  & $\mathrm{M_{\odot} \ yr^{-1}}$\\
DB\_AV & \texttt{Dense Basis} 50th percentile Bayesian A$_V$  & mag\\
DB\_AV\_LOW & \texttt{Dense Basis} 16th percentile Bayesian A$_V$  & mag\\
DB\_AV\_HIGH & \texttt{Dense Basis}  84th percentile Bayesian A$_V$  & mag\\
CI\_MASS & \texttt{CIGALE} 50th percentile Bayesian stellar mass  & $\mathrm{M_{\odot}}$\\
CI\_MASS\_ERR & \texttt{CIGALE} 1$\sigma$ error based on PDF  & $\mathrm{M_{\odot}}$\\
CI\_SFR & \texttt{CIGALE} 50th percentile Bayesian SFR  & $\mathrm{M_{\odot} \ yr^{-1}}$\\
CI\_SFR\_ERR & \texttt{CIGALE} 1$\sigma$ error based on PDF & $\mathrm{M_{\odot} \ yr^{-1}}$\\
CI\_EBV & \texttt{CIGALE} 50th percentile Bayesian E(B-V) & mag\\
CI\_EBV\_ERR & \texttt{CIGALE} 1$\sigma$ error based on PDF  & mag\\
\enddata
\end{deluxetable*}

\subsection{Catalog Depth} 
\label{sec: cat_depth}
We measure the 5$\sigma$ catalog depths for each filter using the final  empirical errors. We  first perform 3$\sigma$ clipping on the errors. We then define the catalog depth as five times the median of that value, and the 1$\sigma$ spread as five times the standard deviation of that value. The point source depths of \cite{finkelstein2025} are 5$\sigma$ limiting magnitudes that were measured in 0\farcs2 diameter circular apertures. Figure \ref{fig: filter_depth} shows the depths for each filter and the values are provided in Table \ref{depths}. 

\begin{deluxetable}{lccc}
\label{depths}
\tablecaption{The median catalog and point source 5$\sigma$ depths for our catalog, as well as the median exposure time. Median catalog depths were measured as described in Section \ref{sec: cat_depth} and the point source 5$\sigma$ depths were measured in 0\farcs2 diameter apertures.}
\tablehead{\colhead{Filter} & \colhead{Median Catalog} & \colhead{Point Source} & \colhead{Median Exp.}\\  
\colhead{} & 
\colhead{5$\sigma$ Depth} & 
\colhead{5$\sigma$ Depth} & 
\colhead{Time (s)}} 
\startdata
F435W & 27.6 & 28.7 & 10423.3 \\
F606W & 27.7 & 28.6 & 3482.1 \\
F814W & 27.5 & 28.3 & 6593.9 \\
F105W & 26.6 & 27.1 & 1935.5 \\
F115W & 28.0 & 29.2 & 5926.7 \\
F125W & 26.9 & 27.3 & 1802.8 \\
F140W & 26.3 & 26.7 & 814.4 \\
F150W & 27.9 & 29.1 & 2834.5 \\
F160W & 27.0 & 27.4 & 3259.3 \\
F200W & 28.2 & 29.3 & 2834.5 \\
F277W & 28.4 & 29.5 & 2834.5 \\
F356W & 28.5 & 29.4 & 3092.2 \\
F410M & 27.8 & 28.7 & 2834.5 \\
F444W & 28.2 & 29.0 & 2834.5
\enddata   
\label{tab: depth}
\end{deluxetable}

\subsection{Flags} \label{sec: flags}

In order to produce a catalog that can be used cover a range of scientific use cases, we opt to include all sources detected by \texttt{SE} in the final catalog, even those that may be spurious. However, we included flags (described below) to indicate to the user sources that may be contaminated by a nearby star, or are otherwise unreliable. We recommend for most users to exclude these flagged sources. In total, there are 87,246 sources in the catalog. After removing flagged objects, 80,755 sources remain. 

\subsubsection{Star Mask} \label{sec: starmask}
In order to indicate which objects might be affected by light from nearby stars or their diffraction spikes, we created a star mask to use as input to \texttt{SE}.

We construct the star mask by identifying stars from the \texttt{SE} segmentation map and the \texttt{SE} cold catalog. We make an initial selection of stars by selecting sources with a FWHM less than 0\farcs1, a CLASS\_STAR value greater than 0.4, and an ELLIPTICITY value less than 0.3. We then manually inspect the results and further refine the list of stars to be masked. 

\begin{figure}
    \centering
    \includegraphics[width=1\linewidth]{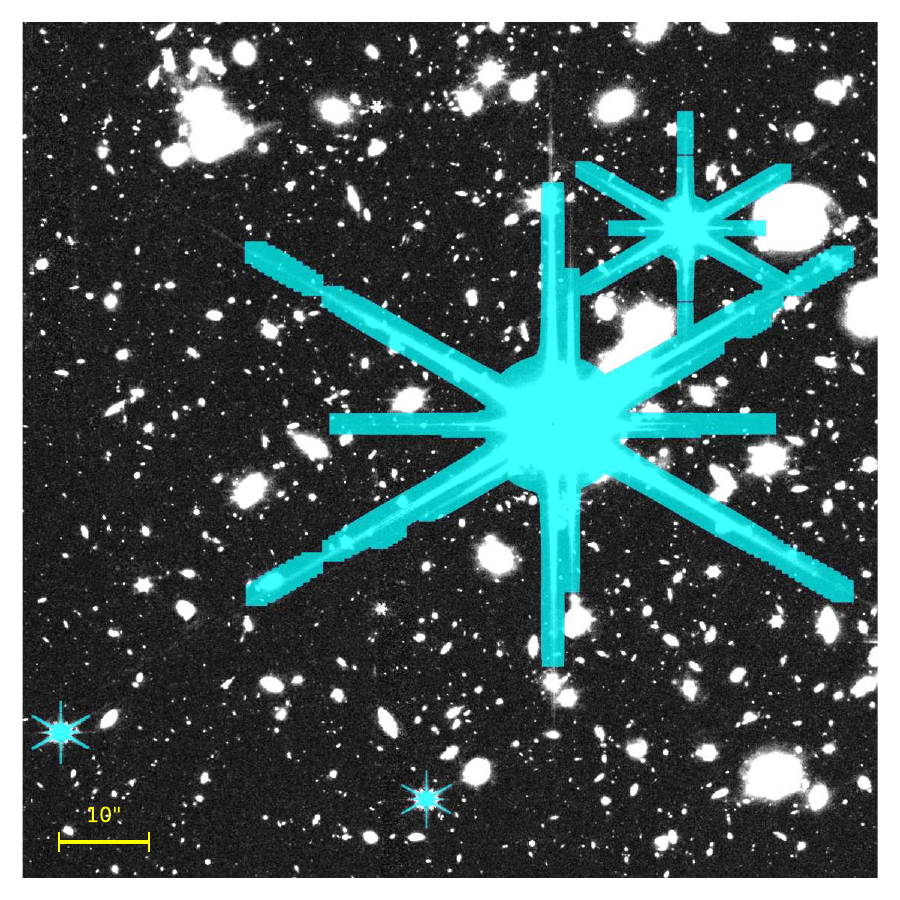}
    \caption{Cutout region of the star mask (magenta) superimposed on top of the detection image. The masks for bright stars were developed as described in Section \ref{sec: flags}.}
    \label{fig: starmask}
\end{figure}

We use a simulated monochromatic PSF that has extended diffraction spikes and added scattered light \footnote{https://www.stsci.edu/jwst/science-planning/proposal-planning-toolbox/simulated-data} and smooth it with a Gaussian kernel. For each star in the list, we construct a rescaled version of the PSF to match the size of the star's FLUX\_RADIUS. We then construct individual star masks by masking regions where the value of the rescaled PSF is above one standard deviation of the background of the detection image mosaic. Finally, we dilate the mask by a factor scaled by the FLUX\_RADIUS of the star to widen the diffraction spikes. A cutout of the star mask around one bright star is shown in Figure \ref{fig: starmask}. Objects that fall within the star mask are denoted by a value of 1 in the STAR\_FLAG column of the catalog.

\subsubsection{Bad Region Mask} \label{sec: bad}

An additional flag is included to indicate sources that may be less reliable or potentially spurious. These sources come from regions of higher noise in the detection image where the short-wavelength filters used in the construction of the detection image do not have complete coverage from the detector gaps or from the F356W-only out of field imaging from the grism exposures.

To identify these regions, we create a total error map where the error images corresponding to each of the filters used in the detection image are added in quadrature and then thresholded to isolate the regions of interest. Whether or not a source lies within the masked region is indicated by a flag in the catalog (BAD\_REGION\_FLAG), and most users will want to exclude sources within the mask as extracted photometric measurements may be less reliable. We denote missing data in the catalog with -99 values. 

\subsection{CEERS Public Data Release 1.0}

The CEERS Public Data Release 1.0 includes the photometric catalog described above, as well as the v1.0 reductions of the NIRCam and MIRI imaging as described in the Appendix of \cite{finkelstein2025}. These data are available for download available for download at \url{ https://ceers.github.io/releases.html} and on MAST as High Level Science Products via doi: \dataset[10.17909/z7p0-8481]{http://dx.doi.org/10.17909/z7p0-8481}.

\section{Summary} \label{sec: summary}

We have presented a deep multi-band photometric catalog based on the seven JWST NIRCam filters from the CEERS Survey and ancillary HST ACS and WFC3 imaging on the same footprint, covering $\sim$90 arcmin$^2$ of the EGS field. All filters were either PSF-matched to our longest wavelength filter (F444W) or, for the case of the HST WFC3 filters, had a PSF-correction applied. We summarize our results as follows:

\begin{itemize}
    \item Photometry was extracted using \texttt{SE}, with a ``hot + cold" approach, where we tuned two separate sets of detection settings to separately prioritize detecting faint and small sources (hot settings) and not over-deblending large extended sources (cold settings). 
    \item Further refinement of the photometry was done by performing aperture corrections, as well as an additional residual aperture correction to account for any light lost in the wings of PSFs. We also estimated robust empirical errors, using an approach that is independent of the errors produced by \texttt{SE}. 
    \item We assessed quality of the photometric catalog by comparing the colors of the available HST bands to the existing CANDELS catalog for the EGS field \citep{stefanon_2017}. We also compared the magnitude differences between our longest wavelength JWST NIRCam imaging (F356W and F444W) to the corresponding IRAC photometry at similar wavelengths (also from the CANDELS catalog). 
    \item We estimated photometric redshifts for the sources in the catalog using \texttt{LePHARE}, and compared them to existing spectroscopic redshifts in the field, finding overall good agreement, with $NMAD$ values of 0.035 at 18$<$F277W$<$23, 0.041 for 23$<$F277W$<$26, and 0.073 at $26>$F277W$<28$ and $\eta$ values of 0.075, 0.09, and 0.176, respectively.    
    \item We estimated physical parameters, including stellar mass, star formation rate, and E(B-V) at the fixed \texttt{LePHARE} redshifts using \texttt{LePHARE}, \texttt{CIGALE}, and \texttt{Dense Basis}, and evaluated the agreement between the codes.
    \item Finally, we release the full catalog, including photometry, photometric redshifts, and stellar parameters along with the full mosaics at each filter as part of the CEERS Public Data Release 1.0. 
\end{itemize}

The catalog is available on the CEERS webpage\footnote{\url{https://ceers.github.io/releases.html}} and we expect that it will support a variety of science goals, including studies of galaxies at a range of redshifts. 

\section{Acknowledgments} \label{sec: acknowledgements}

Support for this work was provided by NASA through grant JWST-ERS-01345.015-A and HST-AR-15802.001-A awarded by the Space Telescope Science Institute, which is operated by the Association of Universities for Research in Astronomy, Inc., under NASA contract NAS 5-26555. This research is based in part on observations made with the NASA/ESA Hubble Space Telescope obtained from the Space Telescope Science Institute, which is operated by the Association of Universities for Research in Astronomy, Inc., under NASA contract NAS 5–26555.

The authors acknowledge Research Computing at the Rochester Institute of Technology for providing computational resources and support that have contributed to the research results reported in this publication. \href{https://doi.org/10.34788/0S3G-QD15}{https://doi.org/10.34788/0S3G-QD15}. The authors also acknowledge the Texas Advanced Computing Center (TACC) at The University of Texas at Austin for providing HPC resources that have contributed to the research results reported within this paper. \href{http://www.tacc.utexas.edu}{http://www.tacc.utexas.edu}

Some of the data presented in this paper are available on the Mikulski Archive for Space Telescopes (MAST) at the Space Telescope Science Institute. The specific observations can be accessed via \dataset[doi:10.17909/z7p0-8481]{https://doi:10.17909/z7p0-8481}.

\clearpage
\begin{appendix}
\section{Point-spread functions} \label{sec:appendix-psf}
\begin{figure}[!h]
    \centering
    \includegraphics[width=0.9\linewidth]{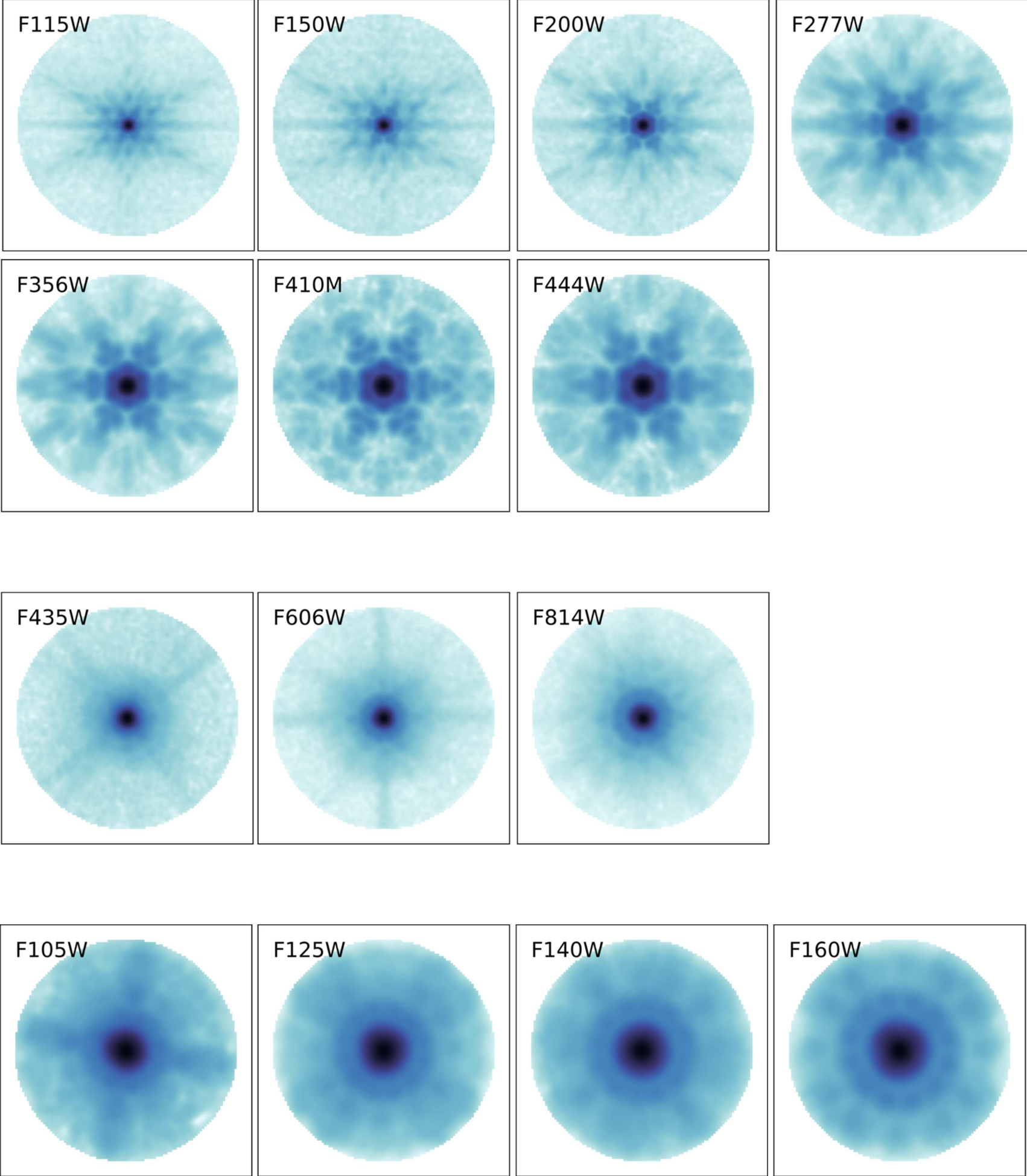}
    \caption{The 14 Point Spread Function (PSFs) we generated using \texttt{PSFEx} with the CEERS imaging. The PSFs are 101x101 pixels with a pixel scale of \farcs03/pixel.}
\end{figure}
\end{appendix}

\clearpage
\bibliography{bib}{}
\bibliographystyle{aasjournal}

\end{document}